\documentclass{aastex63}

\usepackage{xcolor}
\usepackage{soul}
\usepackage{dsfont}
\usepackage{longtable}
\usepackage{graphicx}
\usepackage{amsmath}

\newcommand{\sa}{\langle \sigma \rangle}

\shortauthors{Pirog et al.}
\graphicspath{ {./figures/} }

\begin{document}

\title{Analytical and Numerical Methods for Circumbinary Disk Dynamics - II: Inclined Disks}

\correspondingauthor{Michal Pirog}
\email{mpirog@fau.edu}

\author{Michal Pirog}
\affiliation{The Harriet L. Wilkes Honors College, Florida Atlantic University, Boca Raton, FL 33431, USA}
\affiliation{Department of Physics and Astronomy, West Virginia University, Morgantown, WV 26506, USA
}
\affiliation{Center for Gravitational Waves and Cosmology, West Virginia University, Chestnut Ridge Research Building, Morgantown, WV 26505,
USA
}

\author{Siddharth Mahesh}
\affiliation{Department of Physics and Astronomy, West Virginia University, Morgantown, WV 26506, USA
}
\affiliation{Center for Gravitational Waves and Cosmology, West Virginia University, Chestnut Ridge Research Building, Morgantown, WV 26505,
USA
}

\author{Sean T. McWilliams}
\affiliation{Department of Physics and Astronomy, West Virginia University, Morgantown, WV 26506, USA
}
\affiliation{Center for Gravitational Waves and Cosmology, West Virginia University, Chestnut Ridge Research Building, Morgantown, WV 26505,
USA
}

\begin{abstract}
To gain insight into the dynamical influence of a supermassive black hole binary on a circumbinary accretion disk, we investigate the binary and viscous torque densities throughout such a disk, with emphasis on the final density distribution, particularly the size and stability of the central gap between the binary and the inner edge of the disk. We limit ourselves to the simplified case of a massless, locally isothermal, viscous thin accretion disk under the influence of the gravitational potential from a binary system whose orbital plane is at varying inclinations relative to the disk. In the context of a supermassive black hole binary, the orbital plane could be inclined relative to the circumbinary accretion disk if they are not coeval, or if one or both black holes have spin angular momentum misaligned with respect to the disk's orbital angular momentum, so that the binary can precess to an inclined orientation.  

In our numerical analysis, we employ two-dimensional Newtonian hydrodynamics simulations to examine the influence of two model parameters: the \textit{mass ratio} of the binary and the \textit{inclination angle} between the binary and the disk. Specifically, we investigate their impact on the density and torque distribution. In our analytical approach, we consider the stability of epicycles induced by the perturbative effect of the asymmetric inclined binary gravitational potential on Keplerian circular orbits. We also explore the approximate dynamical torques exerted at resonances.

Through our simulations, we observe that certain configurations never attain a \textit{quasi-steady state}, where the density profile averaged over many orbits stabilizes. This instability occurs when the inclination is close to $45^\circ$, specifically within the range of $\iota \in \langle 35^\circ, 50^\circ \rangle$. It is worth noting that this issue does not arise for \textit{moderately inclined}, \textit{highly inclined}, or \textit{counterrotating} configurations. Furthermore, we identify configurations where there is never a persistent balance between the dynamical and viscous torque densities, as well as cases where the location of this balance oscillates or exhibits other time-dependent behavior over viscous timescales. These findings have implications for understanding both the expected gravitational-wave signal and electromagnetic counterparts from supermassive black hole binaries. 

\end{abstract}

\keywords{black hole binary, disk, torque density, inclination, central gap}

\section{Introduction} \label{sec:intro}

The evolution of gaseous accretion disks surrounding supermassive black hole binaries is a crucial and active area of research in astrophysics. Observationally, understanding the dynamics of these circumbinary disks is vital for interpreting any electromagnetic signatures that might be used to identify supermassive black hole binary candidates \citep{Bartos2017, Stone2017, McKernan2018, Tagawa2020}. Theoretically, the mechanisms by which a gap between the binary and the inner edge of the disk is created and maintained are interesting open problems. Early analytical work on resonant torques and disk dynamics addressing the problem of gap formation in circumbinary disks was presented in \cite{GT1980ApJ, AL96} for coplanar and eccentric cases, respectively. \cite{2015MNRAS.452.2396M} investigated the problem of eccentric binaries inclined with respect to the disk's plane, focusing on the same problem; namely, the balance between binary and viscous torques and its impact on the location of the central gap. Numerically, simulating the progression of the unequilibrated disk towards a quasi-steady state (or, in some cases, the lack thereof) and the maintenance of the circumbinary gap over viscous timescales are significant challenges.

Numerical simulations of these systems have been conducted using both \textit{smooth particle hydrodynamics} (see e.~g.~\cite{AL96, Bate1995, Escala2005, 2007PASJ...59..427H, Cuadra2009, Roedig2012, Pelupessy2013, Ragusa2016}), and \textit{grid-based hydrodynamics} methods (e.~g.~\cite{Gunther, MacFadyen_2008, Hanawa2010, deValBorro2011, DOrazio, Lines2015, Miranda2017}). Our approach to this problem originates from the work of \cite{MacFadyen_2008}, who investigated a two-dimensional coplanar configuration of an equal-mass binary and a thin, viscous, isothermal disk. Since then, several generalizations have been made. \cite{DOrazio} studied different mass ratios of the binary within Newtonian hydrodynamics. Generalizations from pure Newtonian hydrodynamics to Newtonian magnetohydrodynamics (MHD), then to post-Newtonian MHD, and finally to General-Relativistic MHD have been made in e.~g.~\cite{2012ApJ...749..118S, 2012ApJ...755...51N, 2012PhRvL.109v1102F}, respectively. Further studies of the topic can be found, for example, in \cite{Shi2015, Bowen2019}. Additionally, recent studies exploring the vicinity of each black hole and extending the description to the Kerr metric are discussed in \cite{Combi2021, Combi2022}.

Circumbinary disks that are inclined with respect to the binary plane have also been investigated by \cite{2004ApJ...607..913C} in the context of planetary systems. The authors provided evidence that the inclination of the disk may explain the observed electromagnetic flux from the pre-main-sequence star KH 15D. Similarly, \cite{2011PASJ...63..543H} considered the misalignment between circumstellar and circumbinary disks as an explanation for the observed behavior of FS Tauri. In both of these studies, the models' parameters were chosen to match those specific astronomical systems. Analytical studies of disks inclined with respect to the binary plane were carried out by \cite{2014MNRAS.445.1731F}, who derived an analytical description for the long-term evolution of the disk in an axisymmetric perturbed gravitational potential. Also, inclined configurations were investigated further in \cite{Dittmann_from_Sean}.

In this work, we attempt a more general study of the dynamics of circumbinary disks, covering a wide range of our model parameters. However, our approach differs from the aforementioned studies in one important aspect: we do not consider varying eccentricities and instead assume circular orbits. While eccentricity can play an important role, here we specifically concentrate on exploring how the disk dynamics varies with the binary mass ratio and inclination angle. This work is a continuation of our earlier work, \cite{Paper1} (hereafter Paper I), where we studied the same configuration but with zero inclination.

We emphasize that, despite the physical problem being three-dimensional, the hydrodynamical field equations we employ primarily involve two-dimensional functions. The only three-dimensional aspect of this work pertains to the inclination between the binary's orbital plane and the disk, where the total gravitational potential within the disk is due to the combined influence of the two binary components that are executing an inclined circular orbit. In our numerical analysis, we employ two-dimensional Newtonian hydrodynamics simulations to examine the influence of two model parameters: the \textit{mass ratio} of the binary and the \textit{inclination angle} between the binary and the disk. We investigate their impact on the density and torque distribution. It is important to acknowledge the limitations of our two-dimensional simulations. By construction, these simulations do not account for several three-dimensional effects such as disk warping, twisting, disruption, and differential precession, which are likely significant in inclined astrophysical systems \citep{Facchini}. These three-dimensional effects can lead to complex disk morphologies and dynamics that are not captured in our current model.

In disk-binary interactions, it has long been argued that the binary will clear a gap in the disk through resonant torquing \citep{AL96}. In that scenario, the gap size would be determined by imposing that the timescale of angular momentum deposition by the Lindblad Torques (gap-opening timescale) far exceeds the timescale of viscous dissipation of angular momentum through the disk (gap closing timescale) \citep{GT1980ApJ}. Using these ideas, it has been suggested that an electromagnetic afterglow from a circumbinary disk could brighten and be observed long after a pair of massive black holes have merged \citep{2005ApJ...622L..93M}. However, ideas like this rely on the assumption that a gap is maintained due to the resonant behavior of the disk-binary interaction. However, in Paper I, a study of perturbed orbits in the binary potential revealed incredibly high epicyclic advances due to the harmonic modes of the binary potential at a broader range of distances around the corresponding Lindblad resonance. Furthermore, the perturbed orbits at these distances exhibit instabilities with an e-fold timescale much less than the binary orbital period. Thus, the driving of short-timescale instabilities in the inner disk was suggested as being responsible for opening and maintaining circumbinary gaps for these systems.

The effect of misalignment of the binary and disk planes is to weaken the effect of the binary potential \citep{2015MNRAS.452.2396M}. While both gap-opening mechanisms discussed in Paper I rely on the strength of the gravitational potential, the effect of the weakening manifests differently in each picture. The predictions from the orbital stability picture tend to portray a gradual decrease in the size of the circumbinary gap. On the other hand, the weakening of the resonant torques leads to sharp transitions of the dominant Lindblad resonances, sometimes predicting an increase in the gap size close to counter-rotating inclinations. We demonstrate that, while the numerical gap sizes show some transitionary trends, they do not coincide with the inclinations predicted by the resonant torque picture. However, when considering the unstable nature of the quasi-stationary state for these inclinations, the overall trend is still better predicted by the orbital stability picture. 

Astrophysically, misalignment of binary and disk planes can be induced by the GR-induced precession of the spin and orbital angular momenta of the binary over an associated timescale \citep{2014grav.book.....P}. Should the short-timescale picture hold over a range of binary-disk inclinations, the precession timescale would exceed the gap opening timescale. This would allow for an adiabatic treatment of the disk-binary environment where the gap size fluctuates over the range of inclinations accessed by the binary. Additionally, the effect of the inner disk instabilities could contribute to an electromagnetic counterpart to the gravitational radiation from the binary.

In order to compare our analytical predictions with numerical simulations, we utilized the \textit{DISCO} code, which has previously demonstrated its effectiveness as a solver in similar studies (see, e.g., \cite{Farris2014, DOrazio2016, Tang2017, Duffell2020}). A detailed description of this numerical tool can be found in \cite{disco}. Furthermore, an extensive study comparing various numerical codes designed for binary-disk interactions, including \textit{DISCO}, is provided in \cite{duffell2024santa}.

The rest of the paper is structured as follows. Section \ref{sec:tf} provides a detailed description of the model's construction, features, and parameters, accompanied by the necessary mathematical formulas to express it within the framework of Newtonian gravity and hydrodynamics. We then delve into the numerical approach employed, followed by a presentation of our initial data and boundary conditions. In Section \ref{sec:results}, we focus on the results of our numerical simulations as summarized in Table \ref{tab:1} and introduce the concept of the \textit{quasi-steady state}.
For the sake of convenience, we adopt the terms \textit{moderately inclined}, \textit{highly inclined}, and \textit{counterrotating systems} to denote inclinations $\iota \in \left< 0^\circ, 30^\circ \right>$, $\iota \in \left< 55^\circ, 90^\circ \right>$, and $\iota \in \left( 90^\circ, 180^\circ \right>$, respectively. The range of $\iota \in \left( 30^\circ, 55^\circ \right)$ is referred to as the \textit{unstable region}. We present the details of our analytical model in Section \ref{sec:analytic} and compare it with our numerical results in Section \ref{sec:comp}. Finally, Section \ref{sec:conc} provides a summary, draws key conclusions, and outlines potential future research directions.

\section{Theoretical foundations}\label{sec:tf}

\subsection{Analytical preliminaries}\label{sid::theory}

As with our previous study, we work in units where the gravitational constant $G$, the combined mass of the binary $M$, and the semi-major axis of the binary $a$ are all set to unity. With this in mind, we consider the gravitational potential experienced by test particles within the plane of a disk due to a binary system of point masses, such that the binary is inclined with respect to the disk by an angle $\iota$. We again note that we only study the planar motion of the test particles. 

Following \cite{2015MNRAS.452.2396M}, we decompose the gravitational potential in the plane of the disk into azimuthal and temporal harmonics:
\begin{equation}
    \Phi(r,\phi) = -\frac{1}{r} + \sum_{m,n} \Phi_{m,n}(r,\iota )\cos{m\phi - n t},
\end{equation}
where $r$ is the radial distance, $\phi$ is the azimuthal angle within the plane of the disk, $m$ and $n$ are the azimuthal and temporal harmonic indices, and $t$ is the time in units of the binary orbital period.

Compared to the coplanar case studied in Paper I, we note that an additional harmonic index has been introduced. This is to take into account \textit{eccentric} modes, i.e. harmonic modes where the azimuthal frequency is offset from the temporal frequency of the mode due to non-circular behavior. Even when the binary in question is executing a circular orbit, the projection of an inclined orbit onto the plane of the disk is not circular, and therefore allows for eccentric harmonic modes to propagate and affect the disk dynamics.

Using the Wigner matrix formulation from quantum mechanics, we can decompose the gravitational potential within the plane of the disk into spherical harmonic as suggested in
\cite{2015MNRAS.452.2396M} to get the following expression for the Fourier modes $\Phi_{m,n}$:
\begin{equation}
    \Phi_{m,n}(r,q,\iota) = -2\sum_{l = l_\mathrm{min}} \mathcal{Q}_l W_{l,m}W_{l,n}d^l_{n,m}(\iota)r^{-l-1},
\end{equation}
where $l$ labels the index of the spherical harmonic, $\mathcal{Q}_l = (-\mu)^l(1-\mu) + \mu(1-\mu)^l$ is the multipole moment with $\mu$ being the ratio of the secondary mass to the total mass, $W_{l,m}$ is given in Eq.~(9) of \cite{2015MNRAS.452.2396M} and relates to the equatorial spherical harmonics, and $d^l_{n,m}$ is the Wigner ``little'' $d$ matrix \citep{zettili2009quantum}. As with the coplanar case, $l_{\mathrm{min}} = \mathrm{max}(2,m,n)$.

Importantly, we analyze the effect of the Fourier potential as a perturbation of circular orbits in the background Keplerian potential. The effect of the perturbation is to force epicyclic motion about the circular orbits. The amplitude of the epicycles for any Fourier mode is, in this case, given by
\begin{equation}\label{eqn:amplitude_epicycles}
    A_{m,n}(r_0) = \frac{2\frac{\Phi_{m,n}}{r_0}\left(1 - \frac{n}{m\omega}\right) + \partial_r\Phi_{m,n}}{\left| (m\omega - n)^2 - \omega^2 \right|},
\end{equation}
where $\omega = r_0^{-3/2}$ is the Keplerian angular frequency and $r_0$ is the radius of the background circular orbit; see Section 2.2 of Paper I for a detailed derivation of the epicyclic approximation.

In this study, we focus once again on the stability of the epicyclic motion as well as the strength of Lindblad torques in light of the disk-orbit inclination. In comparison to the coplanar case, we notice two additional components --- the time harmonic in the cosine of the Fourier decomposition and the dependence of the strength of the potential (and, correspondingly, the amplitude of epicycles) on the inclination of the binary's orbit. The time harmonic introduces new resonances to the picture which are conventionally referred to as \textit{eccentric resonances}, since the $m \neq n$ modes are only present when the projection of the binary orbital plane onto the disk plane is noncircular. These resonances can be found further away from the circular Lindblad resonances (i.e. when $m = n$), and therefore can occur at orbits further from the typical gap scale where the instability timescale is much larger. On the other hand, the dependence of the epicyclic amplitude on inclination affects the magnitude of the Lyapunov exponents, and this alters the instability timescale. The strength of the potential is weakened for inclined systems, and this brings the gap closer to the binary, which can ultimately be used to explain the observed trend in the gap sizes from our numerical study.

We note here that while the gravitational potential is written in terms of the radius and azimuthal angle, we neglect the role that polar angle or vertical displacement plays in shaping the dynamics of the test particles. This is to ensure that the analytical description matches the numerical setup of the disk hydrodynamics. Ideally, the inclined binary system will force particles to move toward the point masses, displacing the test particles from the disk plane. In the resonant torque picture, a new set of \textit{vertical resonances} \citep{1998ApJ...504..983L} will need to be included in addition to the eccentric resonances discussed earlier. In the orbital stability picture, the dynamics of the test particles will have an additional degree of freedom resulting in an additional Lyapunov exponent that corresponds to instabilities driving the test particles out of the disk plane. The consequence of off-plane dynamics to the formation and maintenance of the circumbinary gap will be left for future studies.

\subsection{Hydrodynamics equations} 
In our setup, the only source of the Newtonian gravitational potential $\Phi(\mathbf{x})$ is two point masses, $m_1$ and $m_2$, moving on fixed circular orbits. The potential at an arbitrary point in three-dimensional space, represented by the vector $\mathbf{x}$, can be expressed as
\begin{equation}
\label{pot}
\Phi = \frac{m_1}{\mathbf{x} - \mathbf{x_1}} + \frac{m_2}{\mathbf{x} - \mathbf{x_2}},
\end{equation}
where $\mathbf{x_1}$ and $\mathbf{x_2}$ are the positions of the point masses.

The dynamics of the massless disk is governed by the standard equations of Newtonian hydrodynamics for mass (areal) density $\sigma(\mathbf{x})$ and fluid velocity $\mathbf{v}(\mathbf{x})$,
\begin{eqnarray}\label{hydro1}
\partial_t \sigma + \nabla \cdot (\sigma \mathbf{v}) & = & 0, \\
\partial_t \mathbf{v} + (\mathbf{v} \cdot \nabla) \mathbf{v} & = & - \frac{1}{\sigma} \nabla p - \nabla \Phi + \mathbf{f}_\nu, \label{hydro2}
\end{eqnarray}
where $p(\mathbf{x})$ represents the pressure, and the viscous force $\mathbf{f}_\nu(\mathbf{x})$ is given by
\begin{equation}
\mathbf{f}_\nu = \nabla \cdot  (\nu \nabla \mathbf{v}) + \nabla \left( \frac{1}{2} \nu \nabla  \cdot \mathbf{v} \right).
\end{equation}
To complete the system of equations and make it solvable, an equation of state and a prescription for the viscosity are also required \citep{accrition_power}.

\subsection{Numerical Setup} \label{num_app}

The numerical calculations are performed using the open-source numerical code \textit{DISCO} \citep{disco}. This code is designed for grid-based, three-dimensional, moving mesh simulations and is specifically tailored to solve Newtonian magnetohydrodynamics problems in axial symmetry. The code's architecture allows for easy reduction of dimensions to handle two-dimensional problems. The moving mesh capability enables the user to optimize accuracy and computational efficiency by selecting the angular velocity of the grid, thereby minimizing diffusive advection errors.

In our study, we consider a three-dimensional configuration consisting of a black hole binary and a thin, viscous, and massless disk inclined with respect to the binary. The first component of our system is the black hole binary, which is modeled using point masses within the Newtonian framework. We define a ``primed'' polar coordinate system, denoted as $(r^\prime, \phi^\prime)$, centered on the center of mass of the binary and coplanar with the binary's orbit. In this coordinate system, the gravitational potential, Eq.~\eqref{pot}, takes the form:
\begin{equation}
\label{pot_cyl}
\Phi( r^\prime ) = \frac{Gm_1}{r^\prime - r^\prime_1} + \frac{Gm_2}{r^\prime - r^\prime_2},
\end{equation}
where $r^\prime_1$ and $r^\prime_2$ are determined by the first model parameter, the \textit{mass ratio} $q = m_1/m_2$ (also denoted $m_1$:$m_2$ in the text); specifically, they are given by $r^\prime_1 = q/(1+q)$ and $r^\prime_2 = 1/(1+q)$.

The second component of our system is the two-dimensional viscous disk. We define a cylindrical coordinate system ($r, \phi, z$) with the origin at the center of the binary, where the surface $z=0$ is tangential to the orbital plane of the binary at an angle $\iota$. In this configuration, the disk exists solely on the surface ($r, \phi, z=0$), and all hydrodynamics quantities are evaluated on this surface. To describe the viscous force, we adopt the \textit{$\alpha$-type viscosity} following the original approach of \citep{1973A&A....24..337S}. Here, the viscosity is given by $\nu(r) = \alpha s^2/r^{3/2}$, where $\alpha$ is a constant, $s(r) = \chi /r^{1/2}$ is the speed of sound, and $\chi = h/r$ represents the constant ratio of the pressure scale height to the radius. For our simulations, we choose $\chi=0.1$.

\begin{figure}
\centering
\includegraphics[width=0.6\linewidth]{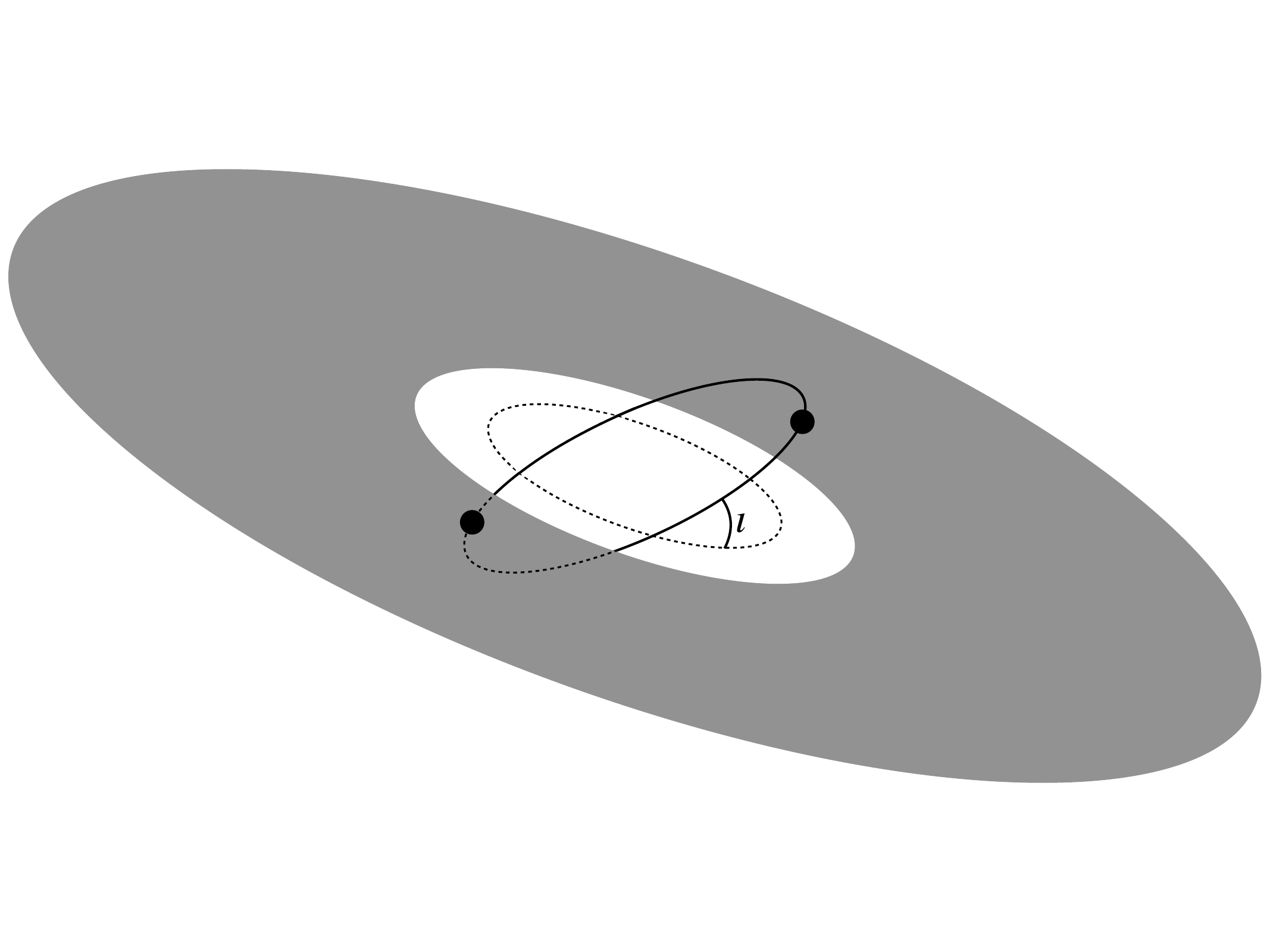}
\caption{The configuration we are considering, where the orbital plane of the binary is inclined by an angle $\iota$ relative to the plane of the disk.}
\label{inclination}
\end{figure}

The system of equations (\ref{hydro1}) and (\ref{hydro2}) consists of four primitive variables: the surface mass density $\sigma$, the pressure $p$, and the two components of the fluid's velocity vector $\mathbf{v} = (v^r, r \Omega)$, all of which depend on the cylindrical coordinates $r$ and $\phi$. Additionally, the pressure is related to the density through the locally isothermal equation of state, given by $p=s^2 \sigma$. Figure \ref{inclination} illustrates the meaning of the \textit{inclination} angle $\iota$ as the angle between the disk and binary planes.

We utilize the 2D version of the open-source Riemann solver, \textit{DISCO} \citep{disco}, specifically the \textit{HLLC} version \citep{toro}, to solve the field equations for the four primitive variables: density $\sigma(t,r,\phi)$, pressure $p(t, r,\phi)$, radial linear velocity $v^r(t, r, \phi)$, and angular velocity $\Omega(t, r,\phi)$. This choice is suitable due to the thinness of the disk. As the point masses are described in the primed coordinates $(r^\prime,\phi^\prime)$, it is necessary to transform their positions to the unprimed coordinates $(r, \phi, z)$ using the following coordinate transformations:
\begin{eqnarray}\nonumber
r_i & = & r^\prime_i \left( \cos^2{\phi_i^\prime}+\sin^2{\phi_i^\prime} \cos^2{\iota} \right)^{1/2}, \\ \label{trans}
\phi_i & = & \arctan{ \left( \tan{\phi_i^\prime} \cos{\iota} \right) }, \\ \nonumber
z_i & = & -r_i^\prime \sin{\phi_i^\prime} \sin{\iota}.
\end{eqnarray}
Here, $i \in \{1,2\}$ corresponds to the first and second mass, respectively. It's important to note that although our model is limited to a 2D problem, we include the 3D information of the point masses' position $(r,\phi,z)$ when calculating the gravitational potential and evaluating the hydrodynamical equations in the 2D setup at the surface $(r, \phi, z=0)$.

\begin{figure}
\includegraphics[width=0.5\linewidth]{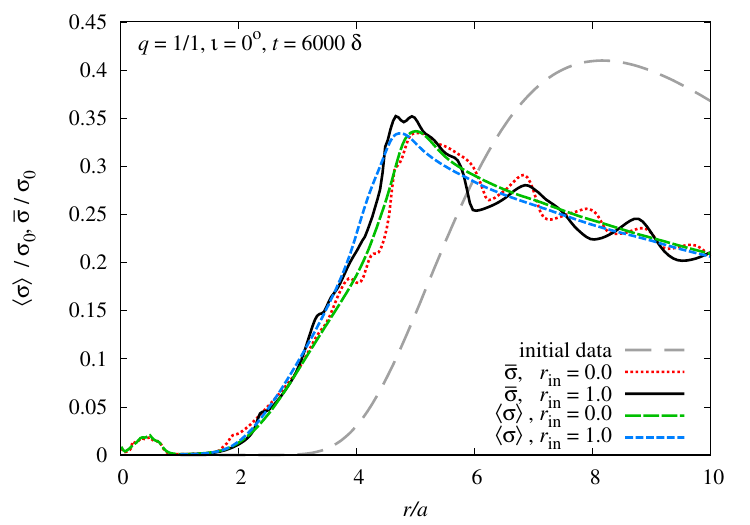}
\includegraphics[width=0.5\linewidth]{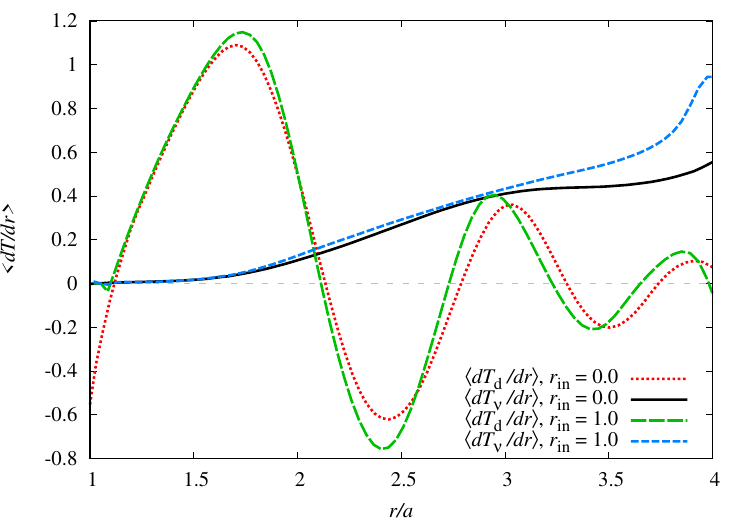}
\caption{The plots show the comparison of our standard approach where we intentionally, due to computational cost, excluded the inner region $r < a$ from our numerical domain (red curves), and a test run without any limits of this kind (green curves). The left plot shows the instantaneous (solid lines) and averaged mass density distributions (dotted lines) for both approaches. The right plot shows that the location where the dynamical (solid lines) and viscous torque densities (dotted lines) intersect does not depend on the choice for the inner boundary. Both plots present a stage of the evolution at $t=6000 \delta$.}
\label{Fig:comp0}
\end{figure}

We adopt the binary separation $a$ as the unit of length. Our computational domain is defined as the region $S=S_\mathrm{100a} \setminus S_\mathrm{a}$, where $S_\mathrm{100a} = r \times \phi : r \le 100a \wedge \phi \in \langle 0, 2\pi)$ is a filled circular region from which we exclude the subdomain $S_\mathrm{a} = r \times \phi : r < a \wedge \phi \in \langle 0, 2\pi)$. The purpose of excluding the inner region $S_\mathrm{a}$ is to optimize computational efficiency, since we are focusing solely on the circumbinary disk. However, it is worth mentioning that future studies will aim to incorporate this excluded region in order to expand the scope of our investigation. For reference, Figure \ref{Fig:comp0} provides a visual comparison between the case labeled as $i00q1$ from Table \ref{tab:1} (equal mass, not inclined) and the corresponding simulation where the input parameters are the same but the region $r < a$ is included in the computational domain.

The numerical grid used in our simulations incorporates moving mesh capabilities, allowing it to rotate with an angular velocity of $\Omega_\mathrm{grid} = r^{-3/2}$. The grid is composed of nearly square-shaped cells, with an aspect ratio close to unity. We introduce the term ``ring'' to refer to a group of grid cells that are equidistant from the center. Our grid consists of a total of $n=480$ radial rings, with each ring containing a varying number of cells denoted by $n_j$, where the index ``$j$'' corresponds to the radial ring. The width of the rings changes with radius, ranging from approximately $0.034$ for the innermost ring to approximately $0.790$ for the outermost ring. Explicitly, the width of the $n$-th ring is given by the difference between two neighboring nodes $\Delta r_n = r_{n+1}-r_n$, where
\begin{equation}\label{node}
r_n = 1 + A \sinh{(BC)},
\end{equation}
with
\begin{equation}\label{nodeex}
A = \frac{5}{\sinh{(1)}}, \quad B = \frac{n-1}{480}, \quad C = \sinh^{-1}{\left(\frac{99}{A}\right)}.
\end{equation}
This particular choice ensures that variations in the density are adequately resolved across multiple cells. The time intervals employed in our simulations vary between time steps and are calculated as half of the minimal propagation time that may occur across the entire grid.
The timescale is expressed in terms of the binary revolution period $\delta$ and is related to the viscous timescale by the well-known approximate formula
\begin{equation}
t_\nu \approx \frac{r^2}{3 \nu} = \frac{1600 \delta}{2^{3/2}} r^{3/2}.
\end{equation}

The majority of our results are expressed in terms of azimuthally and time-averaged quantities. To obtain an exclusively $r$-dependent averaged function $\langle \xi \rangle(r)$ from a general variable $\xi(t, r, \phi)$, we perform the following procedure:
\begin{equation}
\langle \xi \rangle (r)  = \frac{1}{2\pi \Delta} \int^{t+\Delta }_t \int^{2\pi}_0 \xi(t, r,\phi) \ d\phi dt
\label{avg}
\end{equation}
where $\Delta$ is the time interval over which the quantity is averaged. From a numerical perspective, let us introduce the notation $\xi_{ijk}$, which corresponds to the variable $\xi$ evaluated at the $i$th time sample, the $j$th radial ring, and the $k$th azimuthal cell. The averaging procedure can then be written as:
\begin{equation}
\langle{\xi}\rangle_j = \frac{1}{\kappa n_\delta n_j} \sum_{i = m}^{m+\kappa n_\delta} \sum_{k=1}^{n_j} \xi_{i j k}
\label{avg_num}
\end{equation}
where we introduce additional parameters for mathematical correctness and transparency: $\kappa$ represents the number of time samples per binary revolution, $n_\delta$ denotes the number of binary revolutions over which we average, and $m$ represents the number of binary revolutions at which we start the averaging procedure. In our case, we choose $\Delta = 50\delta$, so $n_\delta = 50$ and $\kappa = 60$. This results in 3000 time samples for 50 binary revolutions, which corresponds to one samples for each $6^\circ$ of the binary's circular motion.

\subsection{Initial data and boundary conditions}
Initial data for the four primitive variable is taken from \cite{DOrazio}:
\begin{eqnarray}
\label{id}
\sigma(t=0,r) &=& \sigma_0 \left(\frac{r_\mathrm{s}}{r}\right)^3 \exp{\left(-\frac{r_\mathrm{s}}{r}\right)^2}, \\
p(t=0,r) &=& s^2 \sigma, \\
\Omega(t=0,r) &=& \sqrt{\Omega_\mathrm{K}^2 \left[ 1+\frac{3}{4} \frac{a^2}{r^2} \frac{q}{(q+1)^2} \right]^2 + \frac{\partial_r p}{r\sigma}}, \\
v_r(t=0,r) &=& \frac{2\partial_r (r^3 \nu  \sigma \partial_r \Omega )}{r^2 \Omega \sigma}. 
\end{eqnarray}
Here, $\sigma_0$ and $r_\mathrm{s}$ are constants that determine the value and location of the initial density peak, while $\Omega_\mathrm{K} = r^{-3/2}$ represents the Keplerian angular velocity. It is important to note that this initial data incorporates the mass ratio parameter $q$ but does not include the inclination $\iota$.

The boundary conditions are imposed on the two innermost and outermost radial rings. To apply these conditions, we only need azimuthally averaged quantities:
\begin{equation}
\bar{\xi} (t,r) = \frac{1}{2\pi} \int^{2\pi}_0 \xi(t,r,\phi) \ d\phi.
\end{equation}
In discretized form, this becomes:
\begin{equation}
\bar{\xi}_{ij} = \frac{1}{n_j} \sum_{k=1}^{n_j} \xi_{ij k}.
\end{equation}
The outer boundary conditions are simply the initial data for all primitive variables. For $j \in \{n-1,n\}$, we have:
\begin{eqnarray}
\sigma_{ij} &=& \sigma_{0j} = \sigma_j(t=0), \\ \nonumber
p_{ij} &=& p_{0j} = p_j(t=0),\\ \nonumber
v^r_{ij} &=& v^r_{0j} = v^r_j(t=0),\\ \nonumber
\Omega_{ij} &=& \Omega_{0j} = \Omega_j(t=0).
\end{eqnarray}
On the other hand, the inner boundary conditions for $j \in \{1, 2\}$ are as follows:
\begin{eqnarray}
\sigma_{ij}(t) &=& \bar{\sigma}_{i-1, j+1} (r_{j+1} / r_j)^{-1/2}, \\ \nonumber
p_{ij}(t) &=& \bar{p}_{i-1, j+1} (r_{j+1} / r_j)^{-1/4},\\ \nonumber
v^r_{ij} &=& v^r_{0j} = v^r_j(t=0),\\ \nonumber
\Omega_{ij} &=& \Omega_{0j} = \Omega_j(t=0).
\end{eqnarray}
The index ``i'' corresponds to the number of time steps, where ``i=0'' represents the initial data and ``i-1'' represents the state immediately preceding time step ``i''. Similarly, the index ``j'' corresponds to the number of radial rings, so ``j+1'' corresponds to the next neighboring radial ring. It is important to note that the ``i'' index here denotes time steps dictated by the solving procedure, rather than the number of time samples used for data post-processing. This second order numerical approach requires boundary conditions to be imposed onto two radial rings due to the fact that the Riemann solver uses the cells' \textit{faces}, not nodes or central points. More details about this problem can be found in references \citep{2011ApJS..197...15D, disco}.

\section{Numerical Results}\label{sec:results}

\subsection{Quasi-steady state}

We considered 114 configurations characterized by the viscosity coefficient $\alpha=0.01$, mass ratios $q\in\left\{1\rm{:}1, 2\rm{:}3, 3\rm{:}7, 1\rm{:}4, 1\rm{:}10 \right\}$, and inclinations $\iota \in \left\{0^\circ, 15^\circ, 20^\circ, 25^\circ, 30^\circ, 35^\circ, 40^\circ, 45^\circ, 50^\circ, 55^\circ, 60^\circ, 75^\circ, 90^\circ, 105^\circ, 120^\circ, 135^\circ\right.$, $\left.150^\circ, 165^\circ, 180^\circ \right\}$. They are presented in Table \ref{tab:1}. In addition, we simulated 30 additional configurations with different viscosity coefficients, namely $\alpha=0.03$ and $\alpha=0.003$, ranging over a smaller sampling of mass ratios $q\in\left\{1\rm{:}1, 1\rm{:}4, 1\rm{:}10 \right\}$ and inclinations $\iota \in \left\{0^\circ, 45^\circ, 90^\circ, 135^\circ, 180^\circ \right\}$. These additional configurations are presented in Table \ref{tab:2}, and give us 144 configurations in total.

\startlongtable{
\begin{deluxetable*}{ l c c c c c c c c c c }
\tablecaption{Summary of simulated configurations. All cases have a viscosity coefficient $\alpha=0.01$. The first column contains the run code in the format X-YYY, where X's refer to the mass ratio and YYY encode the inclination angle. The other columns give the mass ratio $q$, inclination angle $\iota$, total simulated time $t_*$, the radius where the viscous and dynamical torque densities balance $r_\mathrm{dT}$ at $t=6000\delta$, the mean of $r_\mathrm{dT}$ over the subsequent 5000 binary revolutions $\tilde{r}_\mathrm{dT}$, the standard deviation of $r_\mathrm{dT}$ over that same time interval $\varsigma(r_\mathrm{dT})$, the radius of the density maximum $r_\mathrm{max}$ at $t=6000\delta$, its mean over the subsequent 5000 binary revolutions $\tilde{r}_\mathrm{max}$, its standard deviation over the subsequent 5000 binary revolutions $\varsigma(r_\mathrm{max})$, and the radius where the density reaches $10\%$ of the final density maximum $r_\mathrm{10\%}$ at $t=6000\delta$. Runs marked with a single asterisk (*) denote those that never reach a quasi-steady state, while those marked with double asterisks (**) denote runs that at first appear to reach a quasi-steady state over many orbits, but eventually depart from the quasi-steady state within a viscous timescale. All other cases persist in their quasi-steady state over a viscous timescale.}
\label{tab:1}
\tablehead{run's code & $q$ &  $\iota$ &       $t_*$ & $r_\mathrm{dT}$ & $\tilde{r}_\mathrm{dT}$ & $\varsigma(r_\mathrm{dT})$ & $r_\mathrm{max}$ & $\tilde{r}_\mathrm{max}$ & $\varsigma(r_\mathrm{max})$  & $r_\mathrm{10\%}$}
\startdata
1-000    & 1:1   & $0^\circ$   & 12000 & 2.0786 & 2.0817 &  0.0045 & 4.7568 & 4.7643 & 0.0342 & 2.3194 \\
1-015    & 1:1   & $15^\circ$  & 12000 & 2.0599 & 2.0749 &  0.0223 & 4.7568 & 4.7719 & 0.0234 & 2.2943 \\
1-020    & 1:1   & $20^\circ$  & 36000 & 2.0632 & 2.0817 &  0.0045 & 4.7568 & 4.7643 & 0.0342 & 2.2922 \\
1-025    & 1:1   & $25^\circ$  & 12000 & 2.1203 & 2.0879 &  0.0572 & 4.7568 & 4.7492 & 0.0186 & 2.2723 \\
1-030    & 1:1   & $30^\circ$  & 30000 & 2.1827 & 2.0665 &  0.0987 & 4.7112 & 4.7492 & 0.0186 & 2.2464 \\
1-035    & 1:1   & $35^\circ$  & 24000 & 2.2467 & 2.0531 &  0.1338 & 4.7112 & 4.7491 & 0.0342 & 2.2163 \\
1-040    & 1:1   & $40^\circ$  & 24000 & 1.9178 & 2.0858 &  0.2130 & 4.7568 & 4.7415 & 0.0371 & 2.1927 \\
1-045*   & 1:1   & $45^\circ$  & 48000 & 1.9681 & 1.8663 &  0.0566 & 4.7568 & 4.8404 & 0.0535 & 2.1935 \\
1-050    & 1:1   & $50^\circ$  & 24000 & 1.7466 & 1.7467 &  0.0001 & 2.4390 & 2.4390 & 0.0000 & 1.7150 \\
1-055    & 1:1   & $55^\circ$  & 12000 & 1.7356 & 1.7357 &  0.0001 & 2.4030 & 2.4030 & 0.0000 & 1.6999 \\
1-060    & 1:1   & $60^\circ$  & 12000 & 1.7240 & 1.7240 &  0.0001 & 2.3672 & 2.3672 & 0.0000 & 1.6647 \\
1-075    & 1:1   & $75^\circ$  & 12000 & 1.6905 & 1.6906 &  0.0002 & 2.2959 & 2.2959 & 0.0000 & 1.5833 \\
1-090    & 1:1   & $90^\circ$  & 12000 & 1.6428 & 1.6427 &  0.0001 & 2.2959 & 2.2959 & 0.0000 & 1.4778 \\
1-105    & 1:1   & $105^\circ$ & 12000 & 1.6029 & 1.6029 &  0.0002 & 2.2959 & 2.2959 & 0.0000 & 1.3683 \\
1-120    & 1:1   & $120^\circ$ & 12000 & 1.5760 & 1.5745 &  0.0013 & 2.5472 & 2.5533 & 0.0148 & 1.2993 \\
1-135    & 1:1   & $135^\circ$ & 12000 & 1.5320 & 1.5282 &  0.0035 & 3.6491 & 3.6557 & 0.0164 & 1.2394 \\
1-150    & 1:1   & $150^\circ$ & 12000 & -      & -      &  -      & 4.1855 & 4.2069 & 0.0234 & 1.2294 \\
1-165    & 1:1   & $165^\circ$ & 12000 & 1.1595 & 1.1595 &  0.0000 & 4.7568 & 4.7794 & 0.0248 & 1.2457 \\
1-180    & 1:1   & $180^\circ$ & 12000 & 1.1617 & 1.1617 &  0.0000 & 4.8479 & 4.8786 & 0.0377 & 1.2576 \\ \hline
2-000    & 2:3   & $0^\circ$   & 12000 & 2.0745 & 2.0736 &  0.0020 & 4.7112 & 4.7340 & 0.0250 & 2.2997  \\
2-015    & 2:3   & $15^\circ$  & 12000 & 2.0913 & 2.0738 &  0.0233 & 4.7112 & 4.7188 & 0.0186 & 2.2788  \\
2-020    & 2:3   & $20^\circ$  & 36000 & 2.0220 & 2.0736 &  0.0020 & 4.7112 & 4.7340 & 0.0250 & 2.2681  \\
2-025    & 2:3   & $25^\circ$  & 12000 & 1.9962 & 2.0550 &  0.0626 & 4.6662 & 4.6887 & 0.0246 & 2.2488  \\
2-030    & 2:3   & $30^\circ$  & 24000 & 2.0295 & 2.0503 &  0.0830 & 4.6662 & 4.6737 & 0.0184 & 2.2313  \\
2-035*   & 2:3   & $35^\circ$  & 24000 & 2.1543 & 2.1149 &  0.1185 & 4.5771 & 4.6439 & 0.0469 & 2.2108  \\
2-040*   & 2:3   & $40^\circ$  & 24000 & 1.8632 & 2.0589 &  0.2060 & 4.6662 & 4.6142 & 0.0593 & 2.1792  \\
2-045    & 2:3   & $45^\circ$  & 48000 & 1.8808 & 1.8414 &  0.0242 & 4.8020 & 4.8021 & 0.0288 & 2.1860  \\
2-050**  & 2:3   & $50^\circ$  & 12000 & 1.7366 & 1.8065 &  0.1280 & 2.4390 & 2.9067 & 0.5343 & 1.6875  \\
2-055    & 2:3   & $55^\circ$  & 12000 & 1.7355 & 1.7352 &  0.0005 & 2.4030 & 2.4090 & 0.0147 & 1.6797  \\
2-060    & 2:3   & $60^\circ$  & 12000 & 1.7214 & 1.7214 &  0.0002 & 2.4030 & 2.4030 & 0.0000 & 1.6574  \\
2-075    & 2:3   & $75^\circ$  & 12000 & 1.6811 & 1.6807 &  0.0004 & 2.2959 & 2.2840 & 0.0184 & 1.5692  \\
2-090    & 2:3   & $90^\circ$  & 12000 & 1.6443 & 1.6442 &  0.0001 & 2.5109 & 2.5109 & 0.0000 & 1.5029  \\
2-105    & 2:3   & $105^\circ$ & 12000 & 1.6062 & 1.6062 &  0.0000 & 2.3315 & 2.3315 & 0.0000 & 1.3884  \\
2-120    & 2:3   & $120^\circ$ & 12000 & 1.5785 & 1.5794 &  0.0008 & 2.9147 & 2.9147 & 0.0000 & 1.2869  \\
2-135    & 2:3   & $135^\circ$ & 12000 & 1.5501 & 1.5467 &  0.0049 & 3.5691 & 3.5757 & 0.0163 & 1.2687  \\
2-150    & 2:3   & $150^\circ$ & 12000 & -      & -      &  -      & 4.0588 & 4.0658 & 0.0171 & 1.2595  \\
2-165    & 2:3   & $165^\circ$ & 12000 & -      & -      &  -      & 4.4883 & 4.4957 & 0.0180 & 1.2526  \\
2-180*   & 2:3   & $180^\circ$ & 12000 & -      & -      &  -      & 4.8020 & 4.8556 & 0.0346 & 1.2581  \\ \hline
3-000    & 3:7   & $0^\circ$   & 12000 & 2.0524 & 2.0521 &  0.0007 & 4.5325 & 4.5325 & 0.0000 & 2.2462  \\ 
3-015    & 3:7   & $15^\circ$  & 12000 & 2.0754 & 2.0458 &  0.0268 & 4.4883 & 4.5104 & 0.0242 & 2.2343  \\
3-020    & 3:7   & $20^\circ$  & 12000 & 2.0595 & 2.0521 &  0.0007 & 4.5325 & 4.5325 & 0.0000 & 2.1915  \\
3-025    & 3:7   & $25^\circ$  & 12000 & 2.0832 & 2.0355 &  0.0639 & 4.4883 & 4.4957 & 0.0180 & 2.1980  \\
3-030    & 3:7   & $30^\circ$  & 24000 & 1.9146 & 2.0224 &  0.1005 & 4.4444 & 4.4518 & 0.0330 & 2.1648  \\
3-035*   & 3:7   & $35^\circ$  & 12000 & 2.1695 & 2.0126 &  0.1275 & 4.3142 & 4.3936 & 0.0508 & 2.1254  \\
3-040    & 3:7   & $40^\circ$  & 12000 & 2.0339 & 2.0111 &  0.1406 & 4.2712 & 4.2855 & 0.0222 & 2.1135  \\
3-045*   & 3:7   & $45^\circ$  & 48000 & 2.0463 & 1.9729 &  0.0427 & 4.4007 & 4.5255 & 0.0857 & 2.1135  \\
3-050    & 3:7   & $50^\circ$  & 12000 & 1.7155 & 1.7288 &  0.0119 & 4.3142 & 4.3502 & 0.0326 & 1.6648  \\
3-055    & 3:7   & $55^\circ$  & 12000 & 1.6951 & 1.6964 &  0.0014 & 4.3142 & 4.3142 & 0.0000 & 1.6560  \\
3-060    & 3:7   & $60^\circ$  & 12000 & 1.6848 & 1.6855 &  0.0005 & 4.2712 & 4.2712 & 0.0000 & 1.6529  \\
3-075    & 3:7   & $75^\circ$  & 12000 & 1.6795 & 1.6795 &  0.0001 & 3.8516 & 3.8516 & 0.0000 & 1.6084  \\
3-090    & 3:7   & $90^\circ$  & 12000 & 1.6358 & 1.6357 &  0.0001 & 3.9339 & 3.9339 & 0.0000 & 1.5234  \\
3-105*   & 3:7   & $105^\circ$ & 12000 & 1.5969 & 1.5970 &  0.0001 & 3.4505 & 3.3852 & 0.0320 & 1.4155  \\
3-120    & 3:7   & $120^\circ$ & 12000 & 1.5811 & 1.5811 &  0.0007 & 3.0651 & 3.0462 & 0.0207 & 1.3067  \\
3-135    & 3:7   & $135^\circ$ & 12000 & 1.5949 & 1.5929 &  0.0038 & 3.3333 & 3.3333 & 0.0000 & 1.3024  \\
3-150    & 3:7   & $150^\circ$ & 12000 & -      & -      &  -      & 3.7702 & 3.7566 & 0.0210 & 1.2841  \\
3-165    & 3:7   & $165^\circ$ & 12000 & -      & -      &  -      & 4.3574 & 4.3790 & 0.0237 & 1.2589  \\
3-180    & 3:7   & $180^\circ$ & 12000 & -      & -      &  -      & 4.8020 & 4.8403 & 0.0345 & 1.2606  \\ \hline
4-000    & 1:4   & $0^\circ$   & 12000 & 1.9934 & 1.9925 &  0.0019 & 4.2283 & 4.2283 & 0.0000 & 2.1530  \\
4-015    & 1:4   & $15^\circ$  & 12000 & 1.9618 & 1.9861 &  0.0260 & 4.1855 & 4.1714 & 0.0218 & 2.1133  \\
4-020*   & 1:4   & $20^\circ$  & 12000 & 2.0120 & 1.9925 &  0.0019 & 4.1008 & 4.2283 & 0.0000 & 2.0770  \\
4-025    & 1:4   & $25^\circ$  & 12000 & 2.0549 & 1.9872 &  0.0580 & 4.0588 & 4.0588 & 0.0000 & 2.0562  \\
4-030    & 1:4   & $30^\circ$  & 24000 & 2.0353 & 1.9567 &  0.0872 & 3.9755 & 3.9962 & 0.0347 & 1.9972  \\
4-035    & 1:4   & $35^\circ$  & 12000 & 2.0981 & 1.9396 &  0.1241 & 3.9339 & 3.9616 & 0.0215 & 1.9682  \\
4-040*   & 1:4   & $40^\circ$  & 12000 & 1.8819 & 1.8603 &  0.0846 & 3.8516 & 3.9002 & 0.1199 & 1.9311  \\
4-045**  & 1:4   & $45^\circ$  & 48000 & 2.0057 & 2.0679 &  0.0573 & 4.0169 & 3.9075 & 0.1533 & 1.9717  \\
4-050    & 1:4   & $50^\circ$  & 12000 & 1.7301 & 1.7240 &  0.0058 & 4.4444 & 4.4737 & 0.0227 & 1.6726  \\
4-055    & 1:4   & $55^\circ$  & 12000 & 1.6733 & 1.6730 &  0.0003 & 4.0169 & 4.0169 & 0.0000 & 1.6312  \\
4-060    & 1:4   & $60^\circ$  & 12000 & 1.6642 & 1.6641 &  0.0004 & 4.0588 & 4.0588 & 0.0000 & 1.6181  \\
4-075    & 1:4   & $75^\circ$  & 12000 & 1.6265 & 1.6269 &  0.0003 & 4.1008 & 4.1008 & 0.0000 & 1.5703  \\
4-090    & 1:4   & $90^\circ$  & 12000 & 1.5962 & 1.5963 &  0.0002 & 4.1432 & 4.1220 & 0.0232 & 1.5011  \\
4-105    & 1:4   & $105^\circ$ & 12000 & 1.5598 & 1.5596 &  0.0001 & 3.9339 & 3.9133 & 0.0226 & 1.4043  \\
4-120    & 1:4   & $120^\circ$ & 12000 & 1.5590 & 1.5595 &  0.0007 & 3.1793 & 3.1665 & 0.0197 & 1.3137  \\
4-135    & 1:4   & $135^\circ$ & 12000 & 1.5950 & 1.5963 &  0.0055 & 3.2175 & 3.2175 & 0.0000 & 1.3163  \\
4-150    & 1:4   & $150^\circ$ & 12000 & -      & -      &  -      & 3.5691 & 3.5558 & 0.0205 & 1.2956  \\
4-165    & 1:4   & $165^\circ$ & 12000 & -      & -      &  -      & 4.2283 & 4.2712 & 0.0272 & 1.2658  \\
4-180    & 1:4   & $180^\circ$ & 12000 & -      & -      &  -      & 4.7568 & 4.8022 & 0.0407 & 1.2659  \\ \hline
5-000    & 1:10  & $0^\circ$   & 12000 & 1.8616 & 1.8616 &  0.0020 & 3.4112 & 3.4440 & 0.0160 & 1.8305  \\ 
5-015    & 1:10  & $15^\circ$  & 12000 & 1.8531 & 1.8329 &  0.0229 & 3.2559 & 3.2817 & 0.0200 & 1.7858  \\
5-020*   & 1:10  & $20^\circ$  & 24000 & 1.8822 & 1.8616 &  0.0020 & 3.3333 & 3.4440 & 0.0160 & 1.7678  \\
5-025    & 1:10  & $25^\circ$  & 12000 & 1.7315 & 1.8123 &  0.0766 & 3.3333 & 3.3398 & 0.0158 & 1.7292  \\
5-030    & 1:10  & $30^\circ$  & 24000 & 1.8619 & 1.7939 &  0.0752 & 3.3722 & 3.3982 & 0.0320 & 1.6829  \\
5-035*   & 1:10  & $35^\circ$  & 24000 & 1.7022 & 1.7042 &  0.0412 & 3.3333 & 3.3918 & 0.0479 & 1.6103  \\
5-040    & 1:10  & $40^\circ$  & 12000 & 1.8393 & 1.8413 &  0.0035 & 3.6090 & 3.5890 & 0.0219 & 1.7081  \\
5-045    & 1:10  & $45^\circ$  & 36000 & 1.6428 & 1.6471 &  0.0098 & 3.6892 & 3.6691 & 0.0220 & 1.4868  \\
5-050    & 1:10  & $50^\circ$  & 12000 & 1.6062 & 1.6064 &  0.0002 & 3.8516 & 3.8516 & 0.0000 & 1.4589  \\
5-055    & 1:10  & $55^\circ$  & 12000 & 1.5969 & 1.5969 &  0.0005 & 3.9755 & 3.9755 & 0.0000 & 1.4531  \\
5-060    & 1:10  & $60^\circ$  & 12000 & 1.5865 & 1.5861 &  0.0004 & 4.0169 & 4.0169 & 0.0000 & 1.4479  \\
5-075    & 1:10  & $75^\circ$  & 12000 & 1.5543 & 1.5535 &  0.0006 & 4.1008 & 4.1220 & 0.0232 & 1.4136  \\
5-090    & 1:10  & $90^\circ$  & 12000 & 1.5225 & 1.5233 &  0.0010 & 4.1008 & 4.1079 & 0.0173 & 1.3652  \\
5-105    & 1:10  & $105^\circ$ & 12000 & 1.4853 & 1.4863 &  0.0008 & 3.6491 & 3.6424 & 0.0163 & 1.2929  \\
5-120    & 1:10  & $120^\circ$ & 12000 & 1.4304 & 1.4332 &  0.0036 & 3.4112 & 3.3852 & 0.0202 & 1.2693  \\
5-135    & 1:10  & $135^\circ$ & 12000 & 1.3766 & 17.5756&  39.6563 & 3.4112 & 3.4112 & 0.0000 &1.2793 \\
5-150    & 1:10  & $150^\circ$ & 12000 & -      & -      &  -       & 3.6491 & 3.6557 & 0.0164 &1.2721 \\
5-165    & 1:10  & $165^\circ$ & 12000 & -      & -      &  -       & 4.1432 & 4.1291 & 0.0219 &1.2547 \\
5-180    & 1:10  & $180^\circ$ & 12000 & -      & -      &  -       & 4.6662 & 4.6963 & 0.0369 &1.2646 \\ \hline
6-000    & 1:100 & $0^\circ$   & 12000 & 1.4017 & 1.4014 &  0.0004 & 3.1030 & 3.1030 & 0.0000 & 1.2789  \\ 
6-015    & 1:100 & $15^\circ$  & 12000 & 1.3709 & 1.3709 &  0.0006 & 2.9147 & 2.8961 & 0.0204 & 1.2450  \\
6-020    & 1:100 & $20^\circ$  & 12000 & 1.3379 & 1.4014 &  0.0004 & 3.1410 & 3.1030 & 0.0000 & 1.2364  \\
6-025    & 1:100 & $25^\circ$  & 12000 & 1.2967 & 1.2908 &  0.0061 & 3.3722 & 3.3398 & 0.0158 & 1.2302  \\
6-030    & 1:100 & $30^\circ$  & 12000 & -      & -      &  -      & 3.4899 & 3.5031 & 0.0203 & 1.2220  \\
6-035    & 1:100 & $35^\circ$  & 12000 & -      & -      &  -      & 3.6090 & 3.6090 & 0.0000 & 1.2172  \\
6-040    & 1:100 & $40^\circ$  & 12000 & -      & -      &  -      & 3.6892 & 3.6959 & 0.0165 & 1.2208  \\
6-045    & 1:100 & $45^\circ$  & 12000 & -      & -      &  -      & 3.7702 & 3.7566 & 0.0210 & 1.2166  \\
6-050    & 1:100 & $50^\circ$  & 12000 & -      & -      &  -      & 3.8106 & 3.8106 & 0.0000 & 1.2226  \\
6-055    & 1:100 & $55^\circ$  & 12000 & -      & -      &  -      & 3.8516 & 3.8448 & 0.0167 & 1.2167  \\
6-060    & 1:100 & $60^\circ$  & 12000 & -      & -      &  -      & 3.8516 & 3.8584 & 0.0168 & 1.2150  \\
6-075    & 1:100 & $75^\circ$  & 12000 & -      & -      &  -      & 3.8926 & 3.9064 & 0.0213 & 1.2136  \\
6-090    & 1:100 & $90^\circ$  & 12000 & -      & -      &  -      & 3.9339 & 3.9408 & 0.0170 & 1.2141  \\
6-105*   & 1:100 & $105^\circ$ & 12000 & -      & -      &  -      & 4.2283 & 4.3071 & 0.0503 & 1.2081  \\
6-120    & 1:100 & $120^\circ$ & 12000 & -      & -      &  -      & 4.3142 & 4.3286 & 0.0223 & 1.2069  \\
6-135    & 1:100 & $135^\circ$ & 12000 & -      & -      &  -      & 4.3142 & 4.3286 & 0.0223 & 1.2066  \\
6-150    & 1:100 & $150^\circ$ & 12000 & -      & -      &  -      & 4.3142 & 4.3430 & 0.0353 & 1.2063  \\
6-165    & 1:100 & $165^\circ$ & 12000 & -      & -      &  -      & 4.3574 & 4.3718 & 0.0223 & 1.2063  \\
6-180    & 1:100 & $180^\circ$ & 12000 & -      & -      &  -      & 4.4007 & 4.4153 & 0.0226 & 1.2183  \\
\enddata
\end{deluxetable*}
}

\startlongtable{
\begin{deluxetable*}{ l c c c c c c c c c c }
\tablecaption{Summary of configurations with different viscosity coefficients. The first column contains the run code now in a Z-X-YYY format, where Z is the viscosity coefficient $\alpha$, and X and YYY remain unchanged from Table \ref{tab:1}. The other columns are the same as in Table \ref{tab:1}.}
\label{tab:2}
\tablehead{run's code & $q$ &  $\iota$ &       $t_*$ & $r_\mathrm{dT}$ & $\tilde{r}_\mathrm{dT}$ & $\varsigma(r_\mathrm{dT})$ & $r_\mathrm{max}$ & $\tilde{r}_\mathrm{max}$ & $\varsigma(r_\mathrm{max})$  & $r_\mathrm{10\%}$}
\startdata
0.03-1-000*   & 1:1   & $0^\circ$    & 6000  & 1.9903 & 1.9764 &  0.0085 & 3.8927 & 3.9133 & 0.0226 & 2.0158 \\
0.03-1-045    & 1:1   & $45^\circ$   & 6000  & 1.8822 & 1.8895 &  0.0071 & 4.3573 & 4.3791 & 0.0238 & 1.8850 \\
0.03-1-090    & 1:1   & $90^\circ$   & 6000  & 1.5862 & 1.5860 &  0.0002 & 2.5471 & 2.5471 & 0.0000 & 1.3653 \\
0.03-1-135    & 1:1   & $135^\circ$  & 6000  & 1.4503 & 1.4496 &  0.0007 & 4.1856 & 4.2282 & 0.0270 & 1.2135 \\
0.03-1-180    & 1:1   & $180^\circ$  & 6000  & -      & -      &  -      & 4.7567 & 4.8023 & 0.0408 & 1.2135 \\ \hline
0.03-4-000*   & 1:4   & $0^\circ$    & 6000  & 1.8991 & 1.8952 &  0.0059 & 3.6089 & 3.6023 & 0.0163 & 1.8324 \\
0.03-4-045    & 1:4   & $45^\circ$   & 6000  & 1.7173 & 1.7158 &  0.0023 & 3.5294 & 3.5228 & 0.0161 & 1.5004 \\
0.03-4-090    & 1:4   & $90^\circ$   & 6000  & 1.5251 & 1.5252 &  0.0001 & 3.9754 & 3.9754 & 0.0000 & 1.3478 \\
0.03-4-135    & 1:4   & $135^\circ$  & 6000  & 1.4337 & 1.4338 &  0.0003 & 3.9340 & 3.9616 & 0.0214 & 1.2204 \\
0.03-4-180    & 1:4   & $180^\circ$  & 6000  & -      & -      &  -      & 4.7114 & 4.7340 & 0.0248 & 1.2204 \\ \hline
0.03-5-000*   & 1:10  & $0^\circ$    & 6000  & 1.7258 & 1.7251 &  0.0081 & 3.2175 & 3.2240 & 0.0289 & 1.5742 \\
0.03-5-045    & 1:10  & $45^\circ$   & 6000  & 1.5568 & 1.5568 &  0.0000 & 3.3723 & 3.3983 & 0.0202 & 1.3103 \\
0.03-5-090    & 1:10  & $90^\circ$   & 6000  & 1.4276 & 1.4276 &  0.0002 & 4.1009 & 4.1009 & 0.0000 & 1.2635 \\
0.03-5-135    & 1:10  & $135^\circ$  & 6000  & 1.3859 & 1.3859 &  0.0000 & 4.0589 & 4.0869 & 0.0217 & 1.2197 \\
0.03-5-180*   & 1:10  & $180^\circ$  & 6000  & -      & -      &  -      & 4.5769 & 4.6290 & 0.0337 & 1.2197 \\ \hline \hline
0.003-1-000   & 1:1   & $0^\circ$    & 24000 & 2.1439 & 2.1537 &  0.0097 & 5.0808 & 5.0965 & 0.0243 & 2.5877 \\
0.003-1-045*  & 1:1   & $45^\circ$   & 24000 & 2.1471 & 1.9636 &  0.1073 & 4.8479 & 4.9875 & 0.1141 & 2.4247 \\
0.003-1-090   & 1:1   & $90^\circ$   & 24000 & 1.7018 & 1.7015 &  0.0007 & 2.3672 & 2.3672 & 0.0000 & 1.6434 \\
0.003-1-135   & 1:1   & $135^\circ$  & 24000 & 1.5761 & 1.5794 &  0.0057 & 2.4030 & 2.4030 & 0.0000 & 1.3441 \\
0.003-1-180   & 1:1   & $180^\circ$  & 24000 & 1.2282 & 1.2334 &  0.0047 & 2.4750 & 2.4869 & 0.0185 & 1.3358 \\ \hline
0.003-4-000   & 1:4   & $0^\circ$    & 24000 & 2.0817 & 2.0819 &  0.0023 & 4.5771 & 4.5696 & 0.0182 & 2.3912 \\
0.003-4-045*  & 1:4   & $45^\circ$   & 24000 & 2.4531 & 2.3322 &  0.1556 & 4.0169 & 3.9552 & 0.1194 & 2.1754 \\
0.003-4-090   & 1:4   & $90^\circ$   & 24000 & 1.5736 & 1.5735 &  0.0014 & 4.6662 & 4.6662 & 0.0000 & 1.7244 \\
0.003-4-135   & 1:4   & $135^\circ$  & 24000 & 1.7151 & 1.7095 &  0.0065 & 2.9897 & 2.9584 & 0.0153 & 1.4500 \\
0.003-4-180   & 1:4   & $180^\circ$  & 24000 & -      & -      &  -      & 2.5472 & 2.5291 & 0.0199 & 1.3357 \\ \hline
0.003-5-000   & 1:10  & $0^\circ$    & 24000 & 1.9492 & 1.9549 &  0.0036 & 3.8106 & 3.8175 & 0.0167 & 2.0753 \\
0.003-5-045*  & 1:10  & $45^\circ$   & 24000 & 1.6155 & 1.7187 &  0.1012 & 4.3142 & 4.2435 & 0.1487 & 1.6343 \\
0.003-5-090   & 1:10  & $90^\circ$   & 24000 & 1.5747 & 1.5713 &  0.0020 & 4.1432 & 4.1362 & 0.0173 & 1.4968 \\
0.003-5-135   & 1:10  & $135^\circ$  & 24000 & 1.6248 & 1.6250 &  0.0045 & 2.9147 & 2.9147 & 0.0000 & 1.4014 \\
0.003-5-180   & 1:10  & $180^\circ$  & 24000 & -      & -      &  -      & 2.4390 & 2.4570 & 0.0197 & 1.3324 \\ \hline
\enddata
\end{deluxetable*}
}

There is no rigorous definition in the literature for the \textit{quasi-steady state}. Our definition is based on the behavior of the averaged density $\sa$, which is computed using equation (\ref{avg_num}). For brevity, we drop the term ``averaged'' and refer to it simply as the ``density'' later in the text. We also drop the index ``$j$'', keeping in mind that the numerical data is discretized for each radial ring. For each configuration, we analyze the state of the system at time intervals spaced by $1000\delta$ ($250\delta$ for $\alpha=0.03$ cases). We introduce here the following notation: $t_m = m \times 1000\delta$ ($m \times 250\delta$ for $\alpha=0.03$ cases) --- the time at which we sample the data, with $m \in \mathds{N}$; $\sa ^\mathrm{max}_m$ --- the maximum value of the density at $t=t_m$; and $r^\mathrm{max}_m$ --- the value of $r$ at which the density reaches its maximum at $t=t_m$.

Unless otherwise noted, we chose $m=6$ to report our results, corresponding to a time $t=6000 \delta$. Additionally, we denote the means of essential quantities over the time samples $m \in \{6, 7, 8, 9, 10, 11\}$ with a tilde symbol. The corresponding standard deviations over that same interval are denoted by $\varsigma()$. Furthermore, we use $t_*$ to represent the total number of binary revolutions over which we investigated each system. The summary of these results is presented in Table \ref{tab:1}.

According to our definition of quasi-steady state, the difference between $r^\mathrm{max}_6$ and the average taken from the locations of the maxima $\tilde{r}^\mathrm{max} = \frac{1}{6}\sum_{i=6}^{11} r^\mathrm{max}_i$ should not exceed the distance between two neighboring grid nodes, denoted as $\epsilon$.

\begin{equation}
\label{cc}
|\tilde{r}^\mathrm{max} - r^\mathrm{max}_6| \leq \epsilon.
\end{equation}

Given that all the maxima are situated between $r=2.0$ and $r=5.0$, the maximum distance between two neighboring nodes, $\epsilon = 0.047$, accounts for less than $3\%$ of the $r$ value. We observed that certain cases, marked with an asterisk or double asterisk in Table \ref{tab:1}, do not conform to our definition of approaching a quasi-steady state.

All cases with $\iota \leq 30^\circ$ and $\iota \geq 55^\circ$, as indicated in Table \ref{tab:1}, satisfy the aforementioned condition (\ref{cc}). Additionally, these cases exhibit density maxima that are either stable or undergo movement/oscillation within one or two zones over 1000 binary revolutions. On the other hand, cases marked with an asterisk in the range of $35^\circ \leq \iota \leq 50^\circ$ demonstrate a relatively stable peak location, although it oscillates over more than two zones throughout the 1000 binary revolutions. Lastly, the cases marked with a double asterisk do not exhibit stable peaks at all, and the density distribution evolves throughout the entire simulation which lasts for multiple viscous timescales at the radii of greatest interest.

Even if the condition (\ref{cc}) is satisfied for some configurations with $m < 6$, we do not consider time samples prior to $m=6$ to ensure that once $\sa ^\mathrm{max}_m$ is reached, the value of the density maximum does not increase. This precaution is taken because although a quasi-steady state can still involve mass loss due to accretion and numerical effects, it should not result in a density buildup, particularly at the peak.

\begin{figure}
\includegraphics[width=\linewidth]{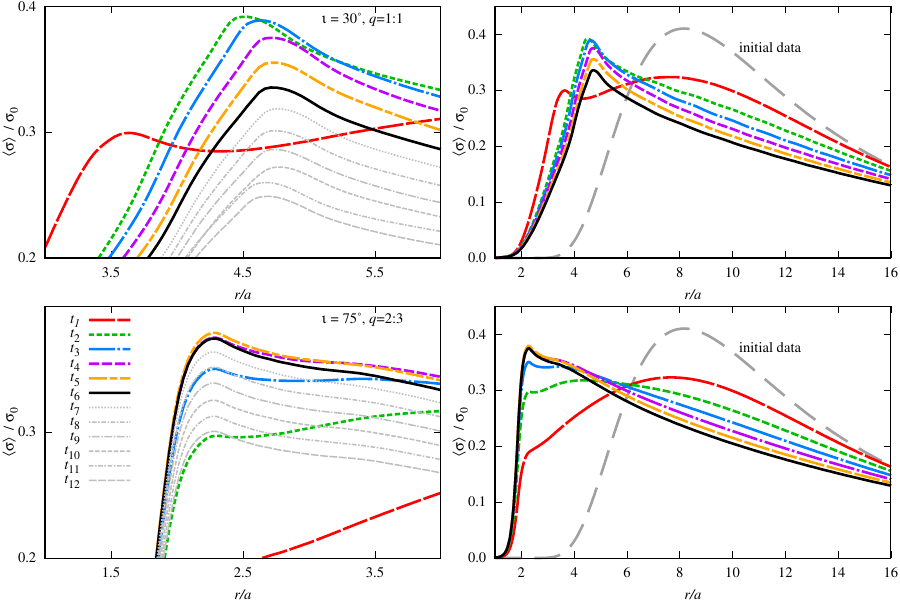}
\caption{The averaged mass density as a function of radius is shown for two arbitrarily chosen cases: $q=1/1$, $\iota=30^\circ$ (upper row) and $q=2/3$, $\iota=75^\circ$ (lower row). In the left panels, the colored lines represent times prior to reaching the quasi-steady state. The final color curve (black) represents the density distribution at the quasi-steady state, specifically at $t=6000\delta$, while the gray lines with different patterns correspond to later stages. The left panels are zoomed in on the regions closer to the peak density, while right panels present the same data over a wider range of radii to show the evolution of the averaged density distribution throughout the disk.}
\label{Fig:steady}
\end{figure}

The two plots on the left-hand side of Figure \ref{Fig:steady} demonstrate the application of the aforementioned definition using examples. The upper left plot depicts the evolution of the maximum and its eventual stabilization at a specific time. Conversely, the lower left plot emphasizes that not only the location of the maximum is significant, but also the value of the density and its temporal evolution. While the location of the maximum is established early on, the matter continues to flow towards the center and accumulate. Furthermore, during the initial stages of the evolution, the maxima are local. For instance, one can look at the blue lines in the right panels for $t_1$ and the dark-blue line for $t_2$ as illustrative examples.

The same single- and double-asterisk notation is applied for the additional cases of different viscosity; however, the time scales differ. In the case of $\alpha=0.03$, we report solutions after $t=3000\delta$, while for $\alpha=0.003$, we report them after $t=12000\delta$.

\subsection{Density distribution}

We analyze the density distributions by focusing on two characteristic radii, denoted $r_\mathrm{max}$ and $r_\mathrm{dT}$. $r_\mathrm{max}$ corresponds to the location of the density maximum. We define $r_\mathrm{dT}$ as the radius where the averaged viscous torque density $\frac{dT_\nu}{dr} = \frac{d}{dr} \left( 2 \pi r^3 \nu  \langle \sigma \frac{d \Omega}{dr} \rangle \right)$
and the averaged dynamical torque density $\frac{dT_\mathrm{d}}{dr} = -2 \pi r \langle \sigma \frac{d \Phi}{d \phi} \rangle$
balance each other:
\begin{equation}
\label{xx}
r_\mathrm{dT} = r: \frac{dT_\mathrm{d}}{dr} = \frac{dT_\nu}{dr}.
\end{equation}
We observe that $r_\mathrm{dT}$ fluctuates/oscillates between time samples, so we report not only the radii $r_\mathrm{dT}$ at $t=t_6$, but also, the average \begin{equation}
\label{x}
\tilde{r}_\mathrm{dT} = \frac{1}{6} \sum_{m=6}^{11} r^m_\mathrm{dT}.
\end{equation}

We present the variations observed in the above radii in two ways. We first present the variation in the mass density distribution for different inclinations at fixed mass ratios. Secondly, we present the variation of the mass density distribution for different mass ratios at fixed inclination angles.
\begin{figure}
\includegraphics[width=\linewidth]{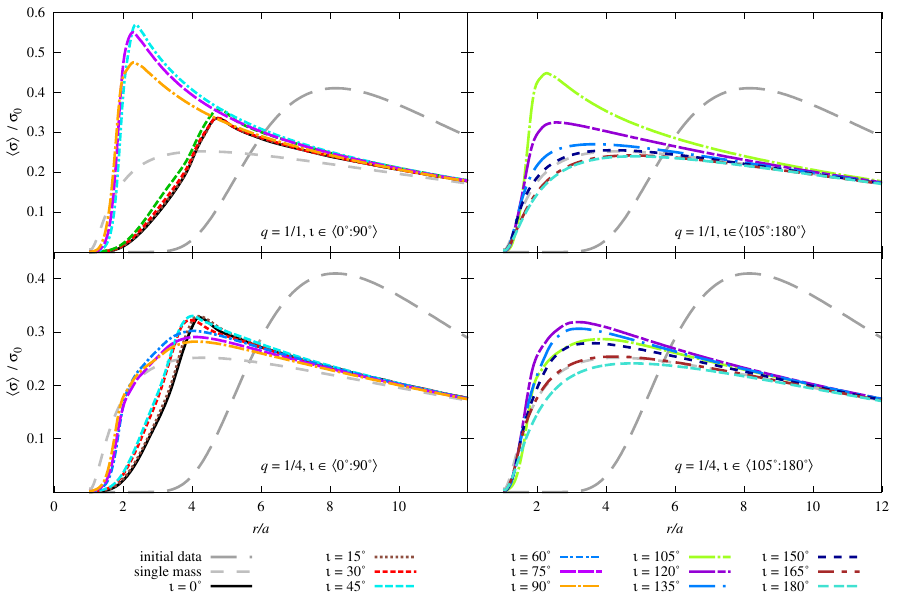}
\caption{Averaged density distribution in the central part of the system for the cases with $q=1$ and $q=1/4$ together with the initial data and single-mass case fore reference.}
\label{dens_a}
\end{figure}

\begin{figure}
\includegraphics[width=\linewidth]{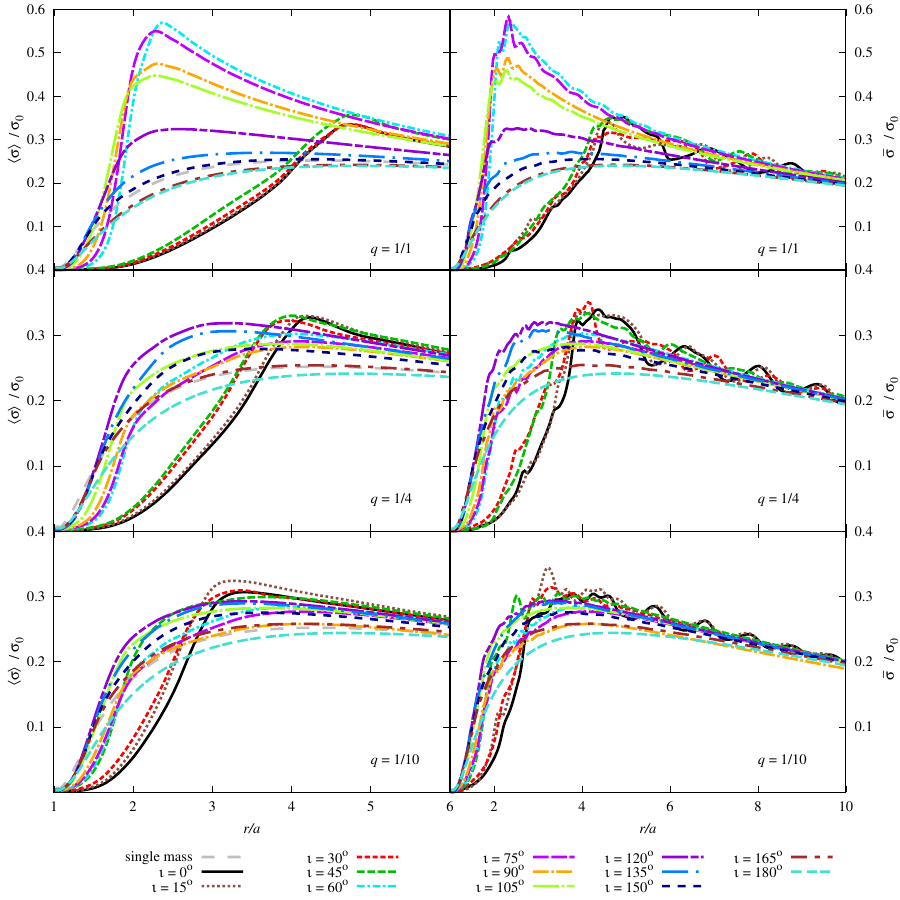}
\caption{Left column: An averaged density distribution in the central part of the system for cases with $q = 1, 1/4, 1/10$. Right column: Instantaneous density (averaged over the azimuthal angle) for the same cases.}
\label{dens_cou}
\end{figure}

The study of the density distribution for $\iota \in \{ 0^\circ, 15^\circ, 30^\circ, 45^\circ, 60^\circ, 75^\circ, 90^\circ,  105^\circ,  120^\circ,  135^\circ,  150^\circ,  165^\circ,  180^\circ \}$ shows an interesting dependence on the inclination angle. Plots in Figure \ref{dens_a} present the locations of $r_\mathrm{max}$ and $r_\mathrm{dT}$, along with the density distributions. They suggest the existence of at least two ranges of inclination angles in which the system behaves in two distinct ways. We first focus on equal-mass cases where the effects are the strongest. In the low inclination sector with $\iota$ less than or equal to $45^\circ$, the density increases moderately with increasing radius, and at a distance of approximately four binary separations from the center, it approaches a maximum value close to the initial data maximum (red, green, dark-violet, and blue curves on the upper left panel). In the case of high inclinations with $\iota$ greater than $45^\circ$, the density has a very steep gradient inside the maximum and grows to values substantially higher than the initial data maximum. In these cases, the locations of $r_\mathrm{max}$ are roughly half of their values for the low inclinations (dark-green, orange, and violet curves on the upper left panel). This simple division into two regimes of inclinations begins to blur, but is still identifiable, as we decrease the mass ratio. In Figure \ref{dens_cou}, on the left, we can see the azimuthally and time-averaged distributions, and on the right, only the azimuthally averaged distribution at a certain time. The observed division strongly manifests in the equal and nearly-equal mass cases, but closer to the extreme case of a binary with a central mass and a satellite, it is significantly weaker but still noticeable. As a matter of fact, we observe intermediate states as we vary the mass ratio from the equal-mass value of $q=1/1$ to the lowest considered value of $q=1/10$.

This division into two domains of behavior is further muddied by the fact that there were configurations, for example 1-045 and 4-045, that were unable to reach a quasi-steady state. Following this observation, we decided to look closer and simulate additional systems with $\iota \in \{ 20^\circ, 25^\circ, 35^\circ, 40^\circ, 50^\circ, 55^\circ \}$. The majority of the additional configurations reached a quasi-steady state according to our requirements, and only some cases, those between $35^\circ \leq \iota \leq 50^\circ$, failed to do so. In principle, larger inclination makes the quasi-steady state more difficult to reach, because the time needed to fulfill condition (\ref{cc}) is getting longer. However, for inclinations at or above $55^\circ$, all configurations are equally well-behaved and clearly reach a quasi-steady state. We present a sample of those density plots in Figure \ref{dens_cou}.

This motivates us to make a more general division where we have \textit{moderately inclined} cases with $\iota \in \langle 0^\circ, 30^\circ \rangle$, \textit{highly inclined} cases with $\iota \in \langle 55^\circ, 90^\circ \rangle$, and, for completeness, \textit{counterrotating} cases with $\iota \in \left( 105^\circ, 180^\circ \right>$. We refer to the subset of cases with $\iota \in \langle 0^\circ, 30^\circ \rangle \cup \langle 55^\circ, 180^\circ \rangle$ as the \textit{stable sector} because they reach the quasi-steady state in a relatively short time. On the other hand, we refer to the configurations with $\iota \in \langle 35^\circ, 50^\circ \rangle$ as the \textit{unstable sector}. Some of these cases do not settle into a quasi-steady state; others have reached this state in the sense of condition (\ref{cc}), but could still potentially become unstable over longer timescales. To explore this possibility, we continued to evolve these cases for a much longer time, up to 48000 binary revolutions (see Tables \ref{tab:1} and \ref{tab:2}), with varying results in terms of stability.

\begin{figure}
\includegraphics[width=\linewidth]{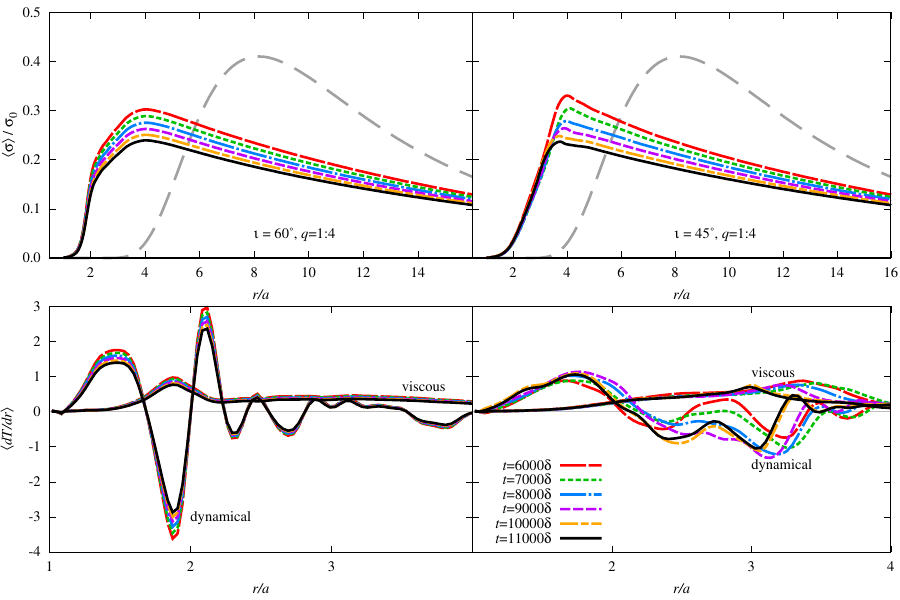}
\caption{Density (upper row) and torque density (lower row) for different time samples for two cases with $q=1/4$: stable one with $\iota=60^\circ$ (left column) and unstable one with $\iota=45^\circ$ (right columns) as an example of what we understand as ``inability to settle down in quasi-steady state''. The color coding is consistent with the color palette in the upper left plot. In the case of torque densities solid and broken lines correspond to dynamical and viscous torques respectively.}
\label{comp}
\end{figure}

To illustrate the dynamics of these two sectors, we compare two cases, both with $q=1/4$, but one with inclination $\iota = 60^\circ$ (in the stable sector) and the other with $\iota = 45^\circ$ (in the unstable sector). The upper-left panel of Figure \ref{comp} confirms the fact that the density peak is not moving for $\iota=60^\circ$. On the other hand, the behavior for $\iota=45^\circ$ presented in the upper-right panel is quite different, with the location of the density peak continuing to evolve steadily. In addition, the torque balance presented in the bottom row reinforces this picture, since the dynamical torque in particular continues to change substantially. The reason for this varying behavior is unclear to us, and will be the subject of future studies.

\begin{figure}
\centering
\includegraphics[width=0.49\linewidth]{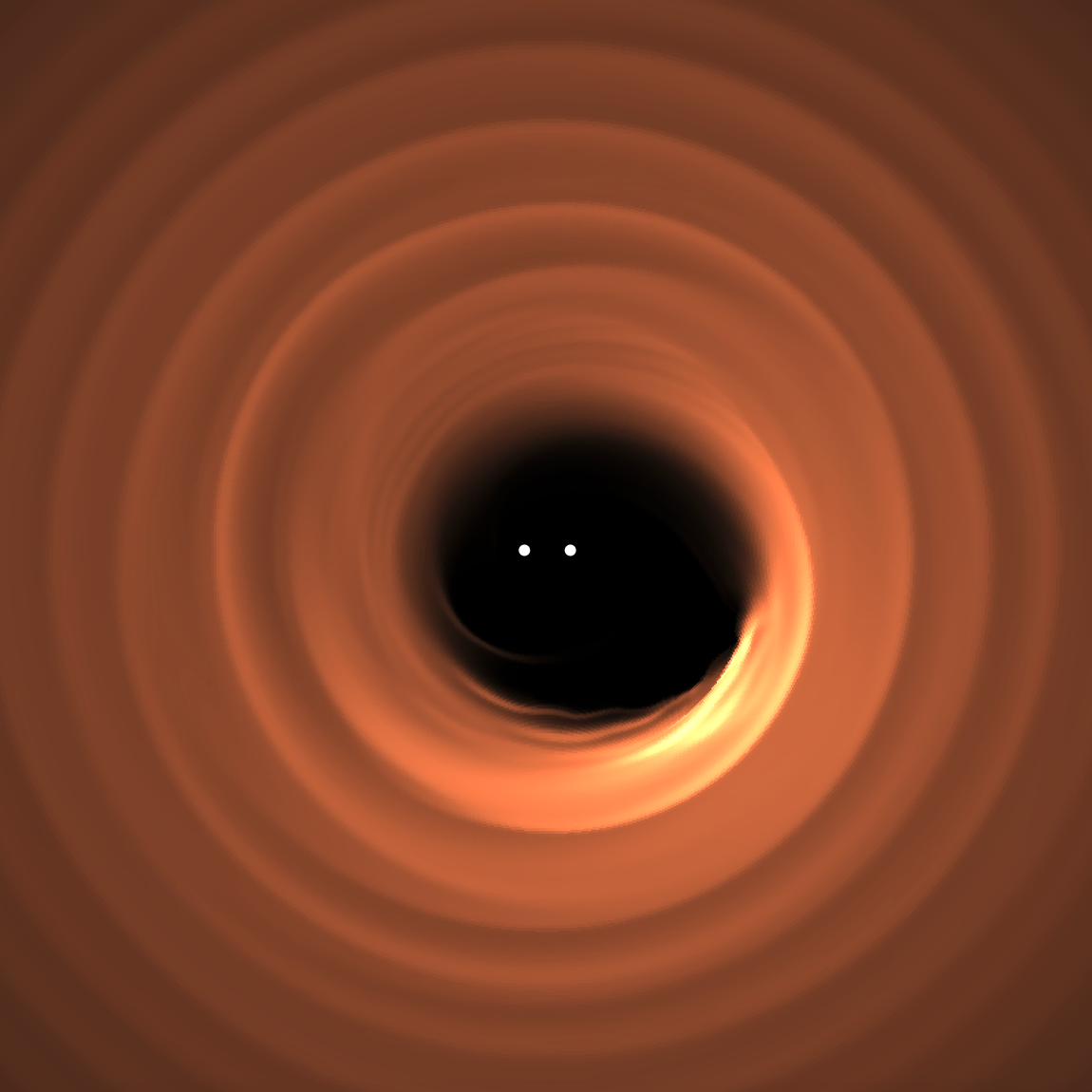}
\includegraphics[width=0.49\linewidth]{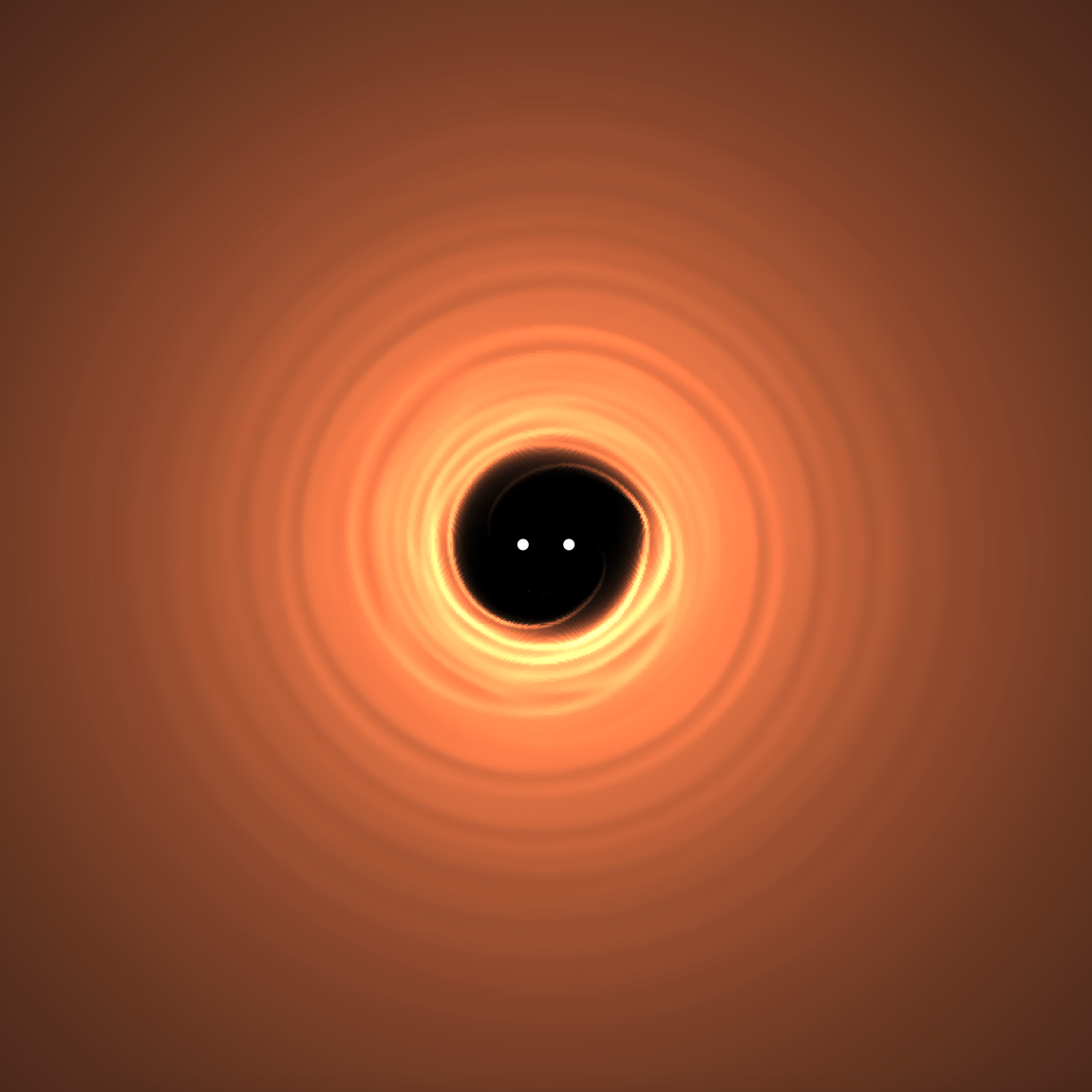}\\
\includegraphics[width=0.49\linewidth]{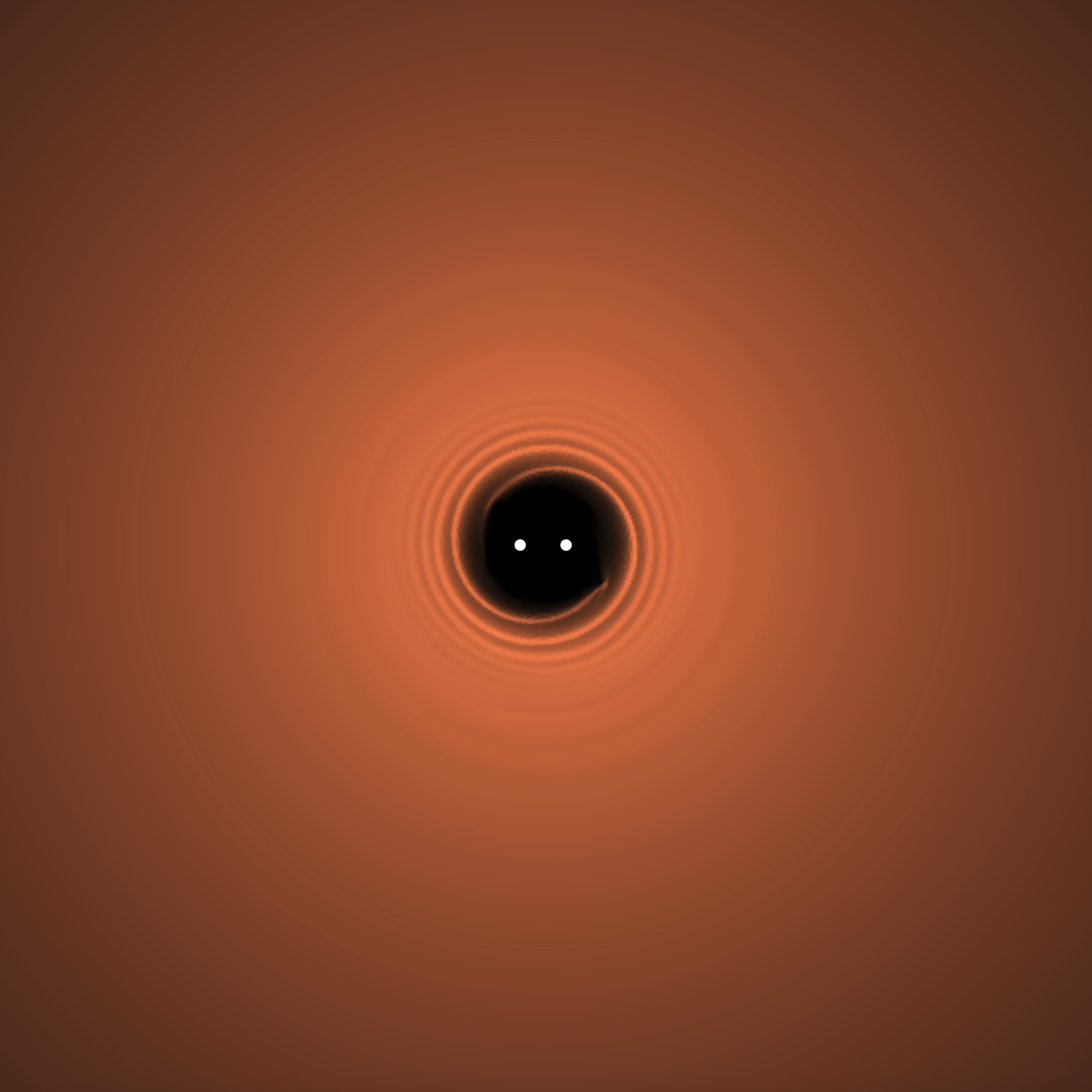}
\includegraphics[width=0.49\linewidth]{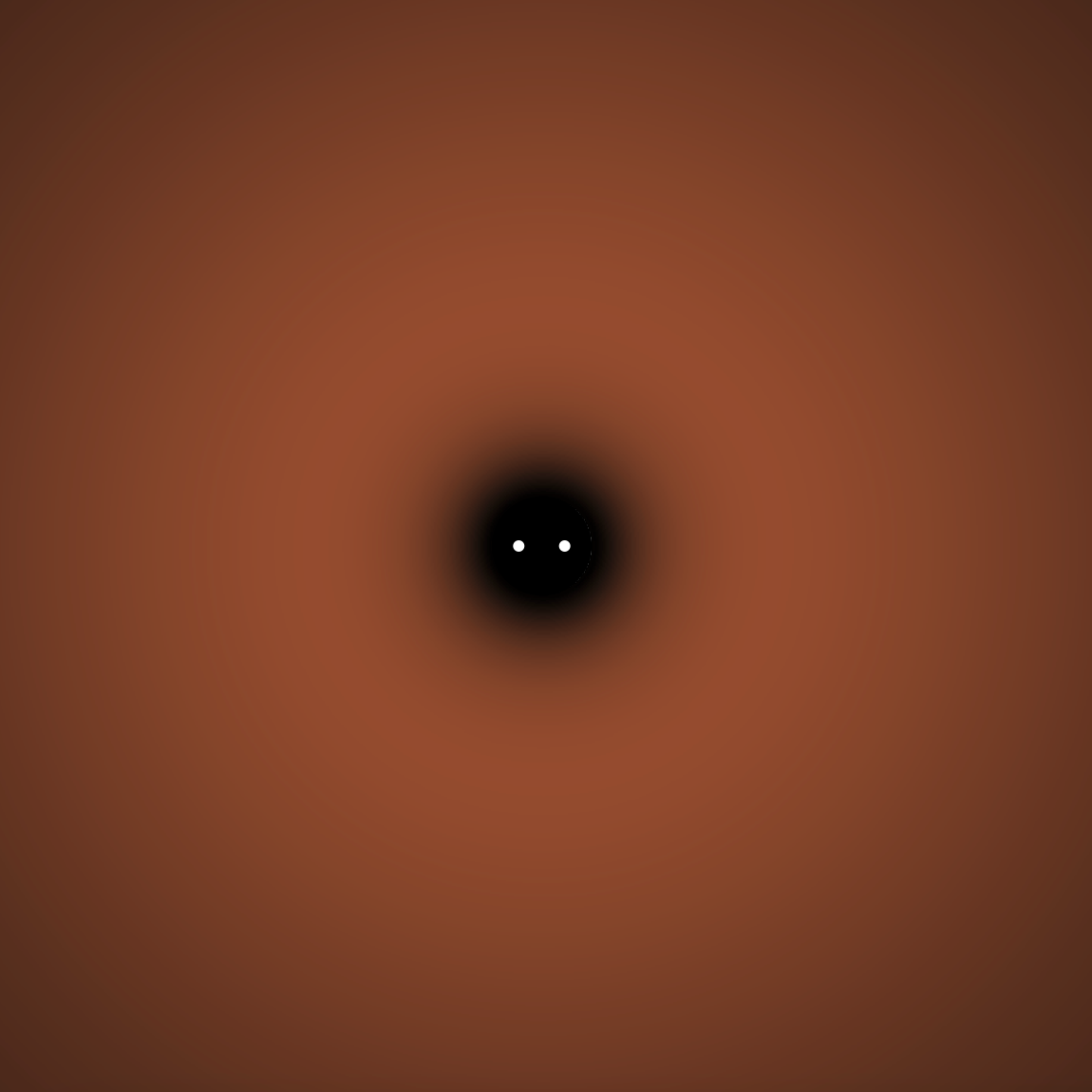}\\
\includegraphics[width=0.986\linewidth]{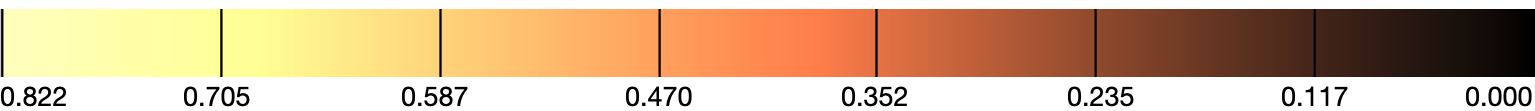}
\caption{Two-dimensional contours of density at $t=6000\delta$ for the equal mass case with $\iota=0^\circ$ (top left), $60^\circ$ (top right), $120^\circ$ (bottom left), and $180^\circ$ (bottom right).}
\label{2dd}
\end{figure}

The differences in the density distribution are noticeable not only in the average density but also in single time samples. We analyzed 2-dimensional snapshots of the density profiles, and found that the division between \textit{not-inclined}/\textit{moderately inclined} and \textit{highly inclined} configurations also manifests in the spatial pattern created by the fluid. For configuration with $\iota \leq 45^\circ$, the density distribution forms the characteristic concentric and closed rings composed of density's local maxima/minima, as in the upper left panel of Figure \ref{2dd}. On the other hand, the \textit{highly inclined} cases,as represented in the upper right panel, form a spiral pattern instead of rings. As we go to still larger inclinations, through the perpendicular and into counterrotating scenario, we see that the matter distribution in the disc becomes increasingly similar to the single mass case, and the effect of the periodic changes in the gravitational potential is not mirrored in the density distribution.

As an additional illustration of these two inclination-dependent regimes, we study the azimuthal mode decomposition of the density distribution,
\begin{eqnarray}
D'_m &=& \frac{1}{2\pi^2} \int_0^{2\pi} d\phi \int_0^{2\pi} d(\Omega_b t) \rho e^{i m (\phi - \Omega_b t)} \quad \text{, if }m\neq0 \\
D_0 &=& \frac{1}{4\pi^2} \int_0^{2\pi} d\phi \int_0^{2\pi} d(\Omega_b t) \rho
\end{eqnarray}
which are numerically calculated as
\begin{equation}
\label{decomposition_num}
D_m = \frac{D'_m}{D_0} = \frac{ \sqrt{ \left[\frac{1}{n_r n_t} \Sigma^{n_t}_{j=0} \Sigma^{n_r}_{i=0} \ \rho_i \cos{(m(\phi_i-t))}\right]^2 + \left[\frac{1}{n_r n_t} \Sigma^{n_t}_{j=0} \Sigma^{n_r}_{i=0} \ \rho_i \sin{(m(\phi_i-t))}\right]^2}} { \frac{1}{n_r} \Sigma^{n_t}_{j=0} \Sigma^{n_r}_{i=0}\rho_i}.
\end{equation}
\begin{figure}
\centering
\includegraphics[width=0.89\textwidth]{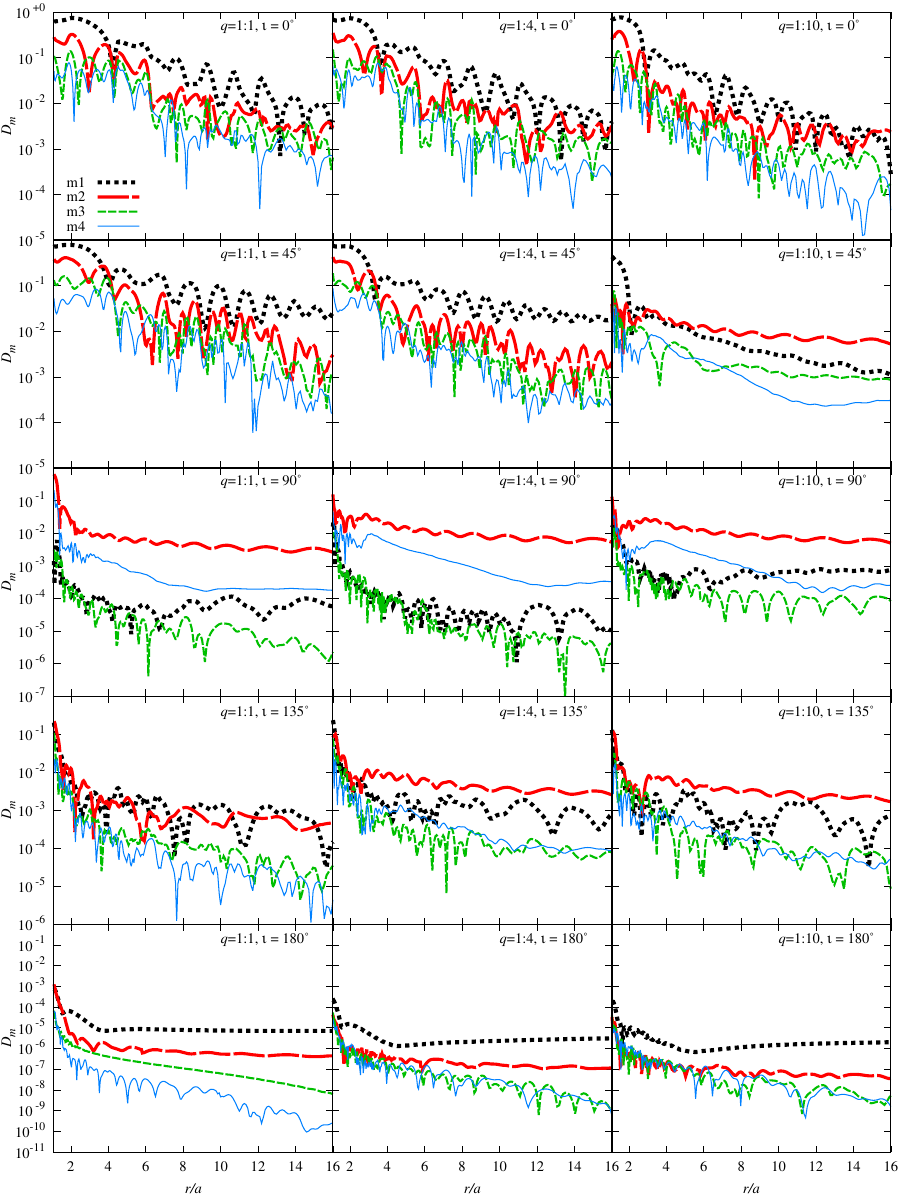}
\caption{Azimuthal mode decomposition for $m \in \{ 1, 2, 3, 4 \} $ corresponding to the density profiles presented in Figure \ref{dens_b}. Columns from the left correspond to mass ratio $q \in \{1/1, 1/4, 1/10\}$ and rows from the top correspond to $\iota \in \{0^\circ, 45^\circ, 90^\circ, 135^\circ, 180^\circ\}$.}
\label{modes}
\end{figure}

We present the results of the azimuthal mode decomposition in Figure \ref{modes}. As in Paper I, we observe that there is no unambiguous decay with increasing mode number beyond $m=1$ at any length scale across the next three $m$ modes. Furthermore, odd-$m$ modes at $q = 1/1$ should be zero since the corresponding harmonic of the gravitational potential is also zero. This is indicative of the fact that, as in the coplanar case, the evolution of the azimuthal modes of the density distribution is not dictated by a linear coupling to the gravitational potential. Therefore, the assertion that there is a dynamical irrelevance of resonant torquing in the opening of the circumbinary gap can be extended to cases where there is an inclination between the disk and binary planes. In our orbital stability picture of gap opening, we sum over multiple modes of the gravitational potential and do not assert that any particular harmonic mode is responsible for the opening of the gap as is the case in the resonant torquing picture.

\begin{figure}
\centering
\includegraphics[width=0.89\textwidth]{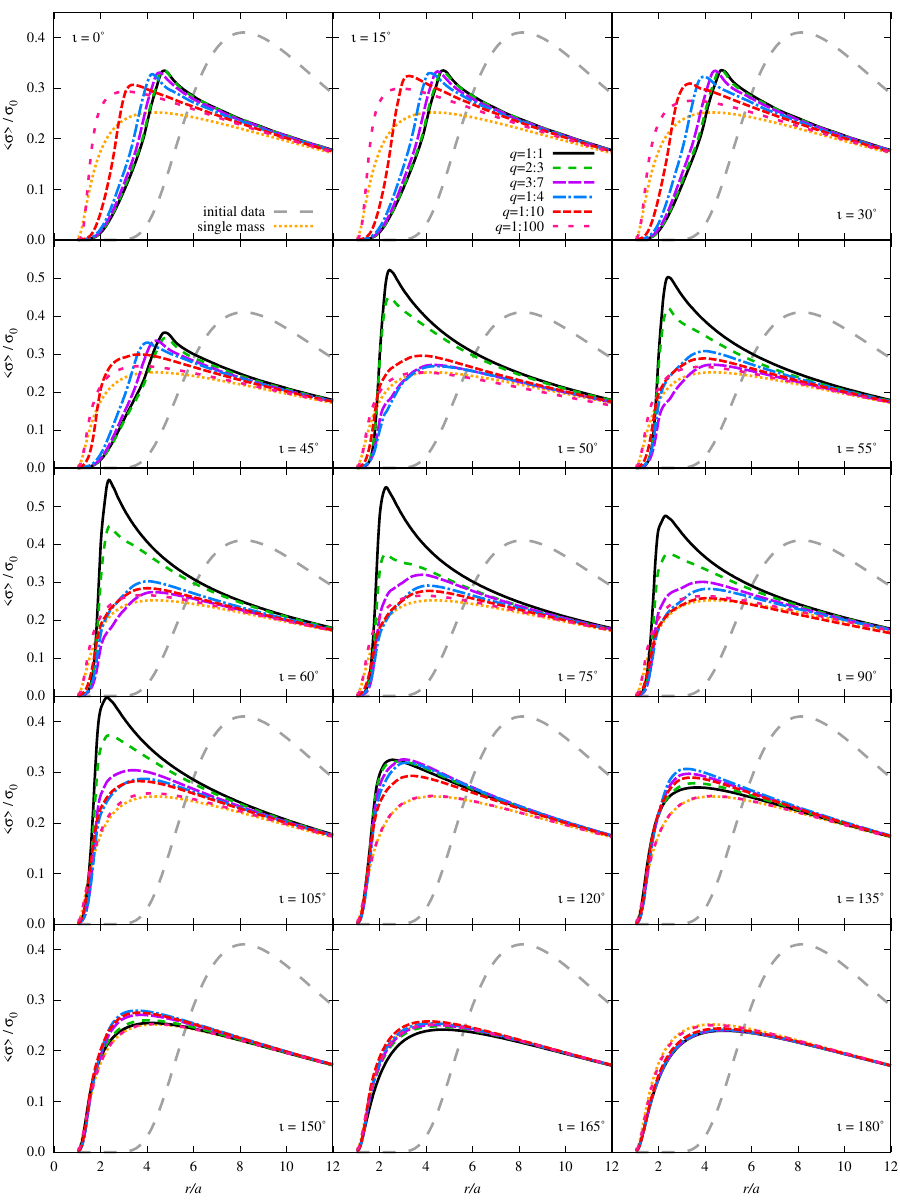}
\caption{Azimuthally and time-averaged density distributions. Each panel shows all mass ratio configurations for a given inclination angle, with panels for all the inclination angles we considered.}
\label{dens_b}
\end{figure}

Another interesting observation is related to mass ratios. In Figure \ref{dens_b}, we compare the density distributions for our full range of mass ratios for a given inclination angle, and provide separate panels for each inclination angle. For  \textit{not-inclined} and \textit{moderately inclined} configurations in the top row, we see a continuous evolution from the equal-mass case down through the lower mass ratios. The situation for larger inclination angles is quite different. From $\iota=50^\circ$ through $105^\circ$, we can see the $q=1$ and $q=2/3$ cases are very different from the other three cases. Interestingly, these three cases, $q \in \{3/7,1/4,1/10\}$, are quite similar to the \textit{highly inclined} systems, which all begin to resemble the single mass case. 

\begin{figure}
\includegraphics[width=\linewidth]{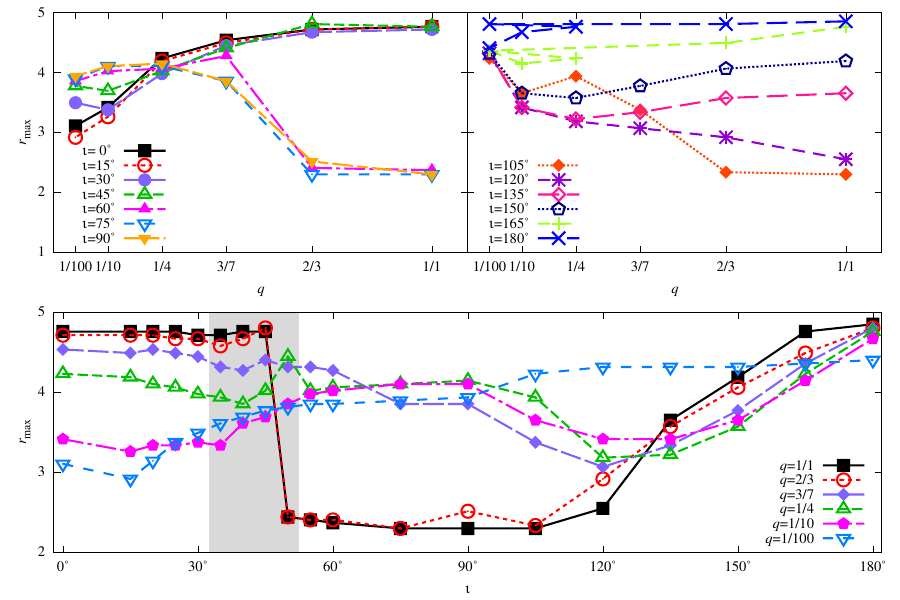}
\caption{The location of $r_\mathrm{max}$. The top panels show different inclination angles as a function of mass ratio, with corotating cases on the left and counterrotating cases on the right (cases $\iota \in \{ 20^\circ, 25^\circ, 35^\circ, 40^\circ, 50^\circ, 55^\circ \}$ are omitted for clarity). The bottom panel shows different mass ratios $q \in \{ 1/1,2/3,3/7,1/4,1/10 \}$ as a function of inclination angle. The gray region indicates the unstable sector.}
\label{Fig:points1}
\end{figure}

\begin{figure}
\includegraphics[width=\linewidth]{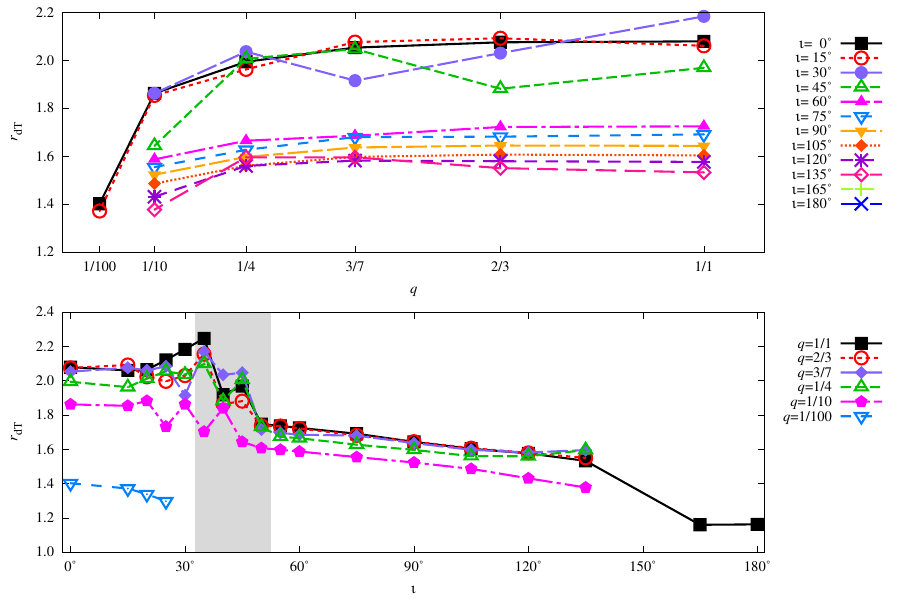}
\caption{The location of $r_\mathrm{dT}$. The top panel shows different inclination angles as a function of mass ratio, but unlike Figure \ref{Fig:points1}, all inclinations are included in a single panel. The bottom panel again shows different mass ratios $q \in \{ 1/1,2/3,3/7,1/4,1/10 \}$ as a function of inclination angle. The gray region indicates the unstable sector.}
\label{Fig:points2}
\end{figure}

\begin{figure}
\includegraphics[width=\linewidth]{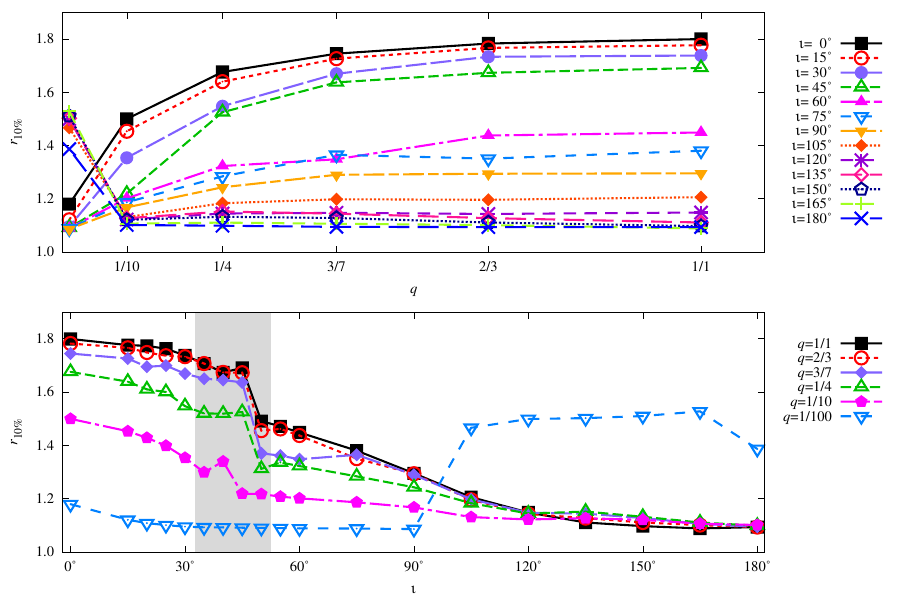}
\caption{The location of $r_{10\%}$. The presentation of different cases is the same as in Figure \ref{Fig:points2}.}
\label{Fig:points3}
\end{figure}

Let's define an additional reference location, the \textit{iso-density radius} $r_{10\%}$, as the radius at which the density has a value of $10\%$ of the final density maximum. 
We now focus on the location of the three radii $r_\mathrm{dT}$, $r_\mathrm{max}$, and $r_{10\%}$, as different ways of characterizing the density distributions more generally. The top panels in Figures \ref{Fig:points1}, \ref{Fig:points2}, and \ref{Fig:points3} present these radii as functions of mass ratio for each inclination angle, while the bottom panels show the radii as functions of inclination angle for each mass ratio. 
We find that the locations of the density maxima for small mass ratio cases $q \in \{3/7,1/4,1/10\}$ vary weakly with inclination angle. On the other hand, for comparable mass cases, i.e $q \in \{1/1,2/3 \}$, the results are completely different from the other three and almost identical with each other --- the location of the density maximum behaves almost like a step function, with the maximum moving sharply inward at high inclinations. For \textit{moderately inclined} cases, the density maxima occur at relatively large radii. For \textit{highly inclined} configurations, with $\iota \geq 55^\circ$, the maxima occur at roughly half the radius of the low inclination cases. 

The trends observed in the behavior of the other two reference radii are less striking and more continuous across mass ratios and inclinations. Both $r_\mathrm{dT}$ and $r_{10\%}$, are large for small inclinations and comparable masses and, with a few exceptions, decrease with decreasing mass ratio. They also decrease with increasing inclination angle, although there are still more exceptions to this monotonicity. All of these exceptions occur in or close to the unstable sector.

\begin{figure}
\includegraphics[width=\linewidth]{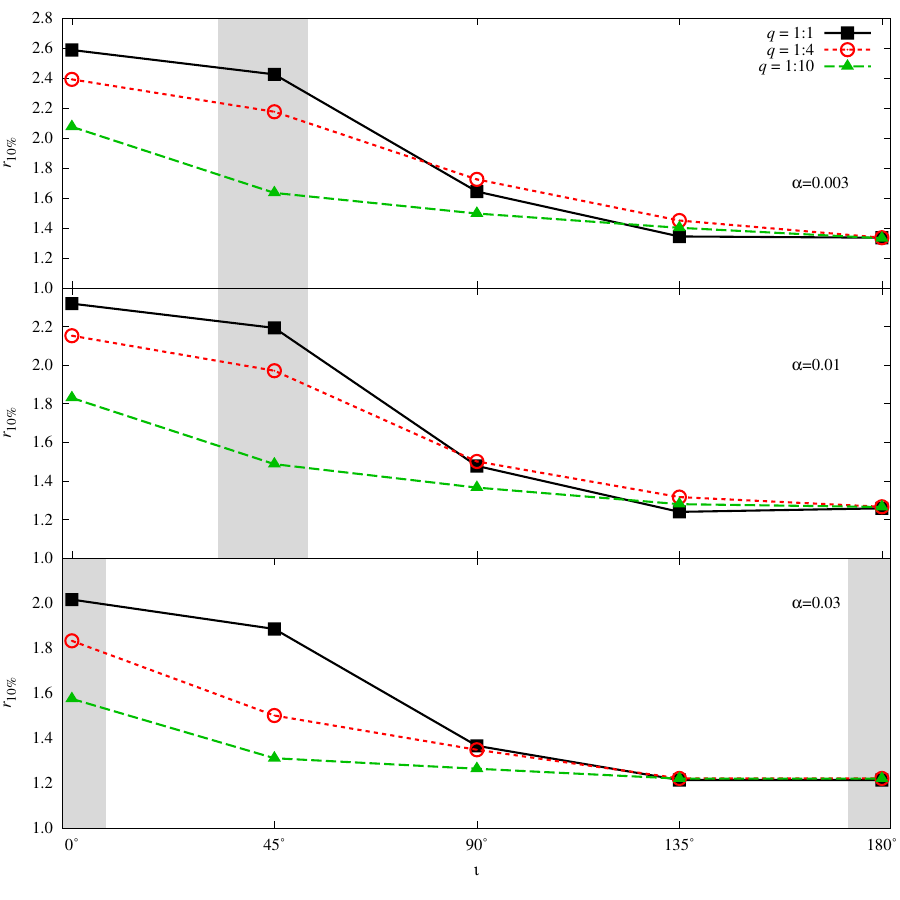}
\caption{The location of $r_{10\%}$ for different inclinations. Each subplot represents a different choice of the viscosity parameter $\alpha$.}
\label{Fig:points4}
\end{figure}

The behavior of the dynamical and viscous torques also bears further study. The bottom row of Figure \ref{comp} displayed how the values of $r_\mathrm{dT}$ stabilize over time. In this paper, we do not focus on the nature or the apparent behavior of the torques, aside from noting that some simulated configurations never reach a consistent value for $r_\mathrm{dT}$. This instability may manifest as completely irregular or secular changes over time, or may oscillate within some restricted range of values. In the second case, the amplitude and period of these oscillations seem to be completely unpredictable and do not depend in any simple way on the model parameters. These effects are present for the entire simulation, long after $t=6000\delta$, which we used to report our data. We pushed some unstable configurations to $24000\delta$, $36000\delta$ or even, in the case of $\iota=45^\circ$, $48000\delta$. Only \textit{not-inclined} or \textit{highly inclined} ($\iota \geq 55^\circ$) systems are free from this effect. 

\begin{figure}
\includegraphics[width=\linewidth]{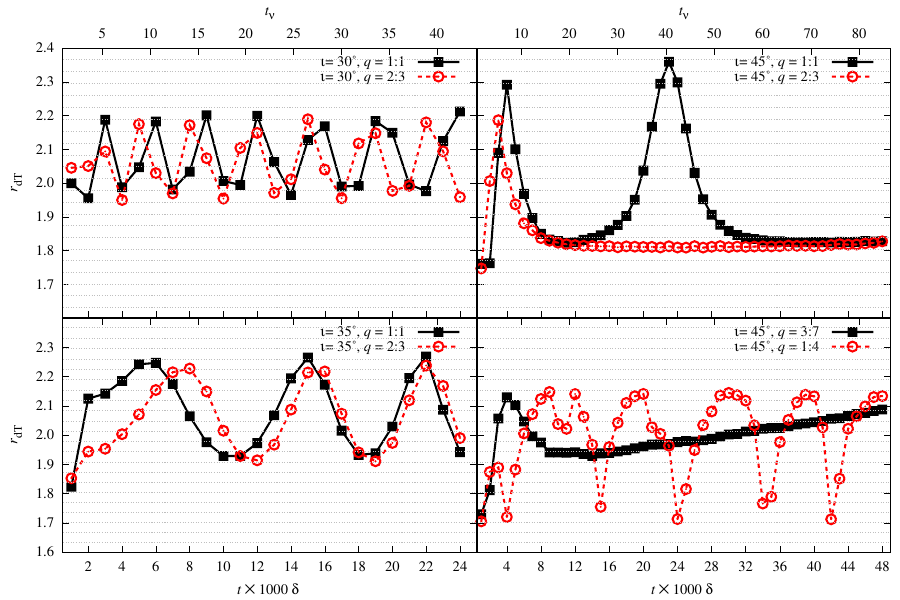}
\caption{The location where the dynamical and viscous torque densities balance, $r_\mathrm{dT}$, for a selection of unstable cases simulated over long times. The gray dotted horizontal lines correspond to the numerical grid's radial nodes.}
\label{Fig:t}
\end{figure}

Figure \ref{Fig:t} illustrates this instability with  some examples of $r_\mathrm{dT}$ calculated at time samples spaced by $1000\delta$. The data here were intentionally chosen to illustrate these qualitatively different behaviors. For inclined cases with $\iota \in \{15^\circ, 20^\circ, 25^\circ, 30^\circ, 35^\circ, 40^\circ \}$ all investigated systems presented some periodicity. We observed that the amplitude was largest and the period was longest for $\iota=30^\circ$.

The inclination $\iota=45^\circ$ behaved qualitatively differently and requires separate discussion. In the case of $\iota=45^\circ$, we observed periodicity for $q=1/4$, stability for $q=2/3$, and neither stability nor periodicity for $q=1/1$ and $q=3/7$.

Lastly, we must emphasize that our conclusions regarding the physical mechanism that dominates the gap and our observations regarding instabilities may depend on the range of viscosities under consideration. In Figure \ref{Fig:points4}, we show $r_{10\%}$ as a function of inclination for $q \in \{1\rm{:}1,1\rm{:}4,1\rm{:}10\}$, for three different viscosities $\alpha \in \{0.003,0.01,0.03\}$. While the general trend of decreasing gap size with increasing inclination is universal, the gap shrinks with increased viscosity at lower inclinations, and is independent of viscosity at larger inclinations. In addition, the unstable sector that we have discussed appears to apply for lower viscosity, but is qualitatively different for the larger viscosity case, where only the not-inclined and counterrotating cases manifest instability. We leave the study of larger viscosity cases to future work.

\section{Analytical Calculations}\label{sec:analytic}

In this section, as with Paper I, we compute the size of the circumbinary gap under two regimes. The first, which is valid at small timescales, relates to the stability of the epicyclic orbits and defines the gap size as the radius where orbital instabilities propagate at timescales equal to the binary orbital period. This has proven fruitful in explaining the observed trends in the gap sizes in the coplanar case from numerical simulations. The second method that assumes the gap is maintained by the balancing of resonant and fluid torques defines the gap size as the radius of the resonance where the corresponding torque densities balance. 

\subsection{Orbital stability and gap size}

To recap from Paper I, we study orbital stability of the epicyclic orbits using Lyapunov exponents. Each Lyapunov exponent corresponds to a particular eigenmode through which a perturbation from an exact solution to the Hamilton's equations of motion propagates. The eigenvector provides the dynamical nature of the perturbation while the eigenvalue, known as the Lyapunov exponent, provides the inverse timescale over which the perturbation evolves exponentially. That is, it is a complex number whose real part determines the inverse timescale over which the perturbation decays or blows up while the imaginary part provides the frequency of the oscillatory component of the perturbation.

In this case, we are interested in the leading or largest Lyapunov exponent as that corresponds to the instability that blows up over the shortest timescale. We then define the gap radius prediction due to orbital instabilities as the radius, $r_\mathrm{L}$, where the following condition is met
\begin{equation} \label{rl}
    \mathfrak{Re}\left(\lambda^{\mathrm{max}}(r_\mathrm{L})\right) = \frac{1}{\delta} = \frac{1}{2\pi}.
\end{equation}
Here, $\lambda^{\mathrm{max}}$ is the largest Lyapunov exponent and $\delta = 2\pi$ is the binary orbital period in our units of choice.

\begin{figure}
\centering
\includegraphics{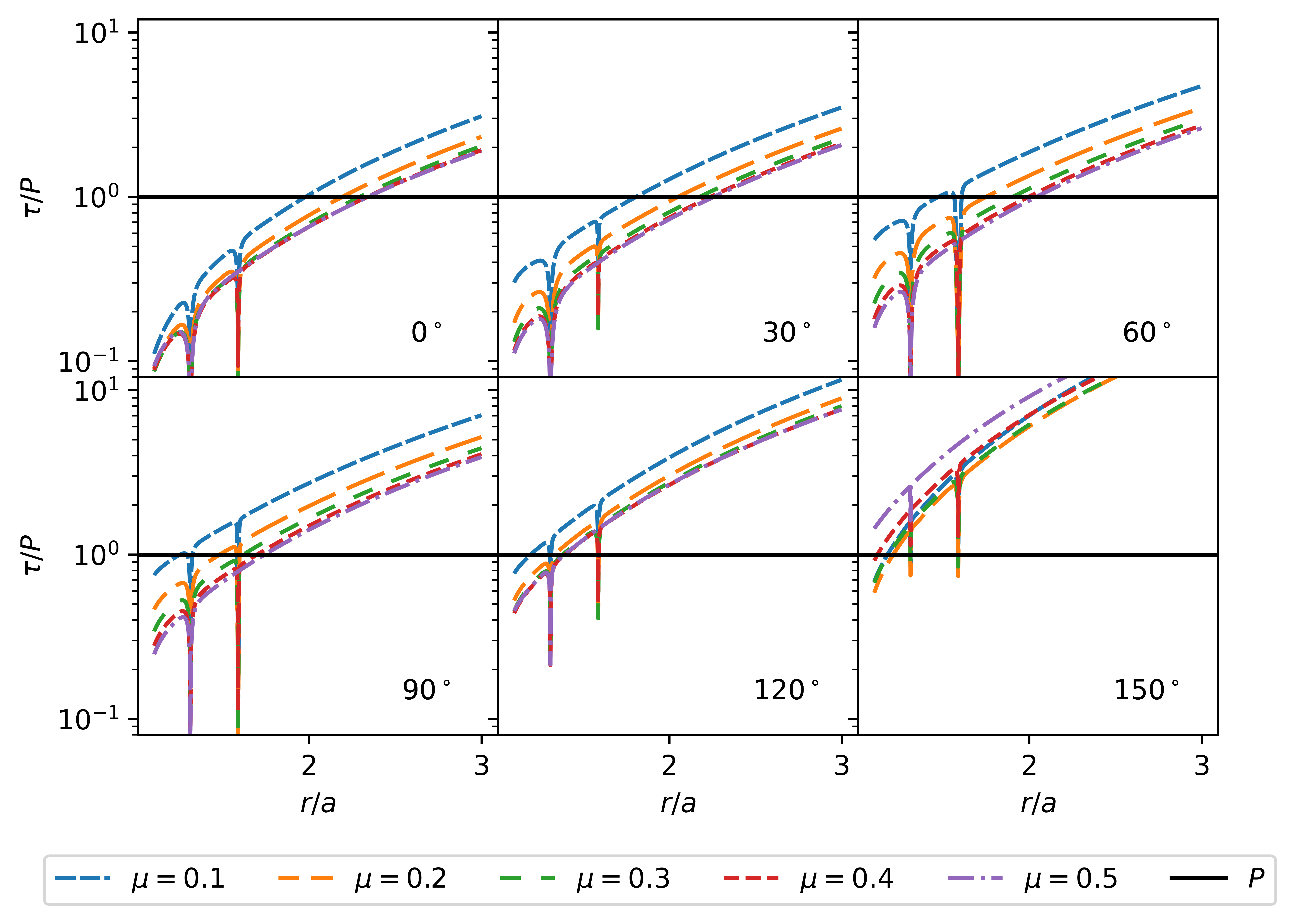}
\caption{The timescale associated with the largest Lyapunov exponent for all mass ratios at inclinations $\iota \in \{15^\circ, 30^\circ, 45^\circ, 60^\circ, 75^\circ, 90^\circ\}$. The timescales have been normalized to the binary orbital period as opposed to our choice of units where the binary frequency is set to unity. The black line corresponds to the binary orbital period. }
\label{Fig:instabilities}
\end{figure}

We present results for $\tau = 1/\mathfrak{Re}(\lambda^{\rm max})$, the instability timescale corresponding to the largest Lyapunov exponent, in Figure \ref{Fig:instabilities}. We note that the instability timescale at each inclination increases with decreasing mass ratio and increasing inclination. This is a result of the weakening of the perturbing potential in those limits. As we approach the counterrotating case, the instability timescales become larger than $\delta$.

\subsection{Resonant torquing picture}

To again briefly recap from Paper I, we study the dynamics of the disk in the fluid picture where the gap is created by the balance of torques applied by the gravitational force of the binary as well as the torque due to viscous dissipation. An analytical description of the torques is based on the WKB approximation of the hydrodynamic equations, where angular momentum is deposited into the disk by dynamical torques at Lindblad resonances and dissipated away by viscous torques. The circumbinary gap is said to be opened at the outermost Lindblad resonance where the dynamical torque is greater than the viscous torque. 
In this description, we first determine the outermost Lindblad resonance where the resonant torque and the viscous torque can balance each other, referred to as the gap opening resonance. Upon determining the gap opening resonance, we define the size of the gap,  $r_\mathrm{T}$, as the outermost location where the amplitude of the epicycles, $A(r_\mathrm{T})$, given by Eq.~\eqref{eqn:amplitude_epicycles}, can extend to the location of the gap opening resonance, $r_{\zeta}$. That is, we solve the equation

\begin{equation}\label{eq:rT}
    f(r_\mathrm{T}) \equiv \log\left(\frac{A(r_\mathrm{T})}{|r_\mathrm{T} - r_\zeta|}\right) = 0.
\end{equation}

\begin{figure}
\centering
\includegraphics{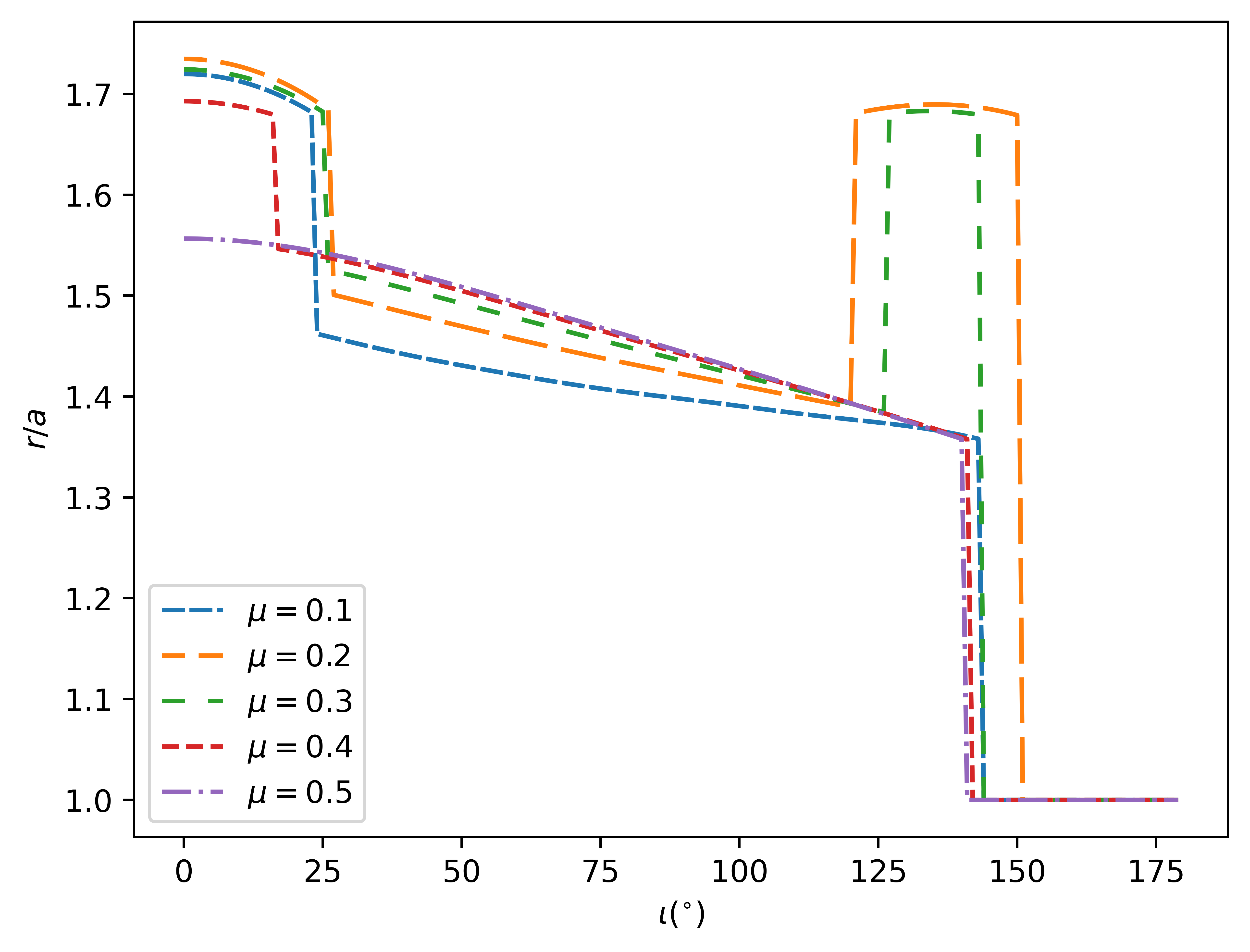}
\caption{The value of $r_\mathrm{T}$ given by the balancing of the resonant and viscous torques in the WKB approximation.}
\label{Fig:torquegapsizes}
\end{figure}

We present the results of the gap size according to the resonant torquing picture in Figure \ref{Fig:torquegapsizes}. We note that, as with Paper I, the equal mass case yields a gap size smaller than other mass ratios since the symmetry in masses destroys the $(m,n)=(1,1)$ resonance. Additionally, we note that at each mass ratio, there is a critical inclination at which the gap opening resonance transitions from $(1,1)$ to $(2,2)$. This is due to the weakening of the gravitational potential with increasing inclination. Finally, we note that when approaching the counterrotating case, the gap size shows an increase due to the weakening of the resonant torque at these inclinations leading to non-circular ($m \neq n$) gap opening resonances.

\section{Comparison of analytical and numerical results}\label{sec:comp}

\begin{figure}
\centering
\includegraphics{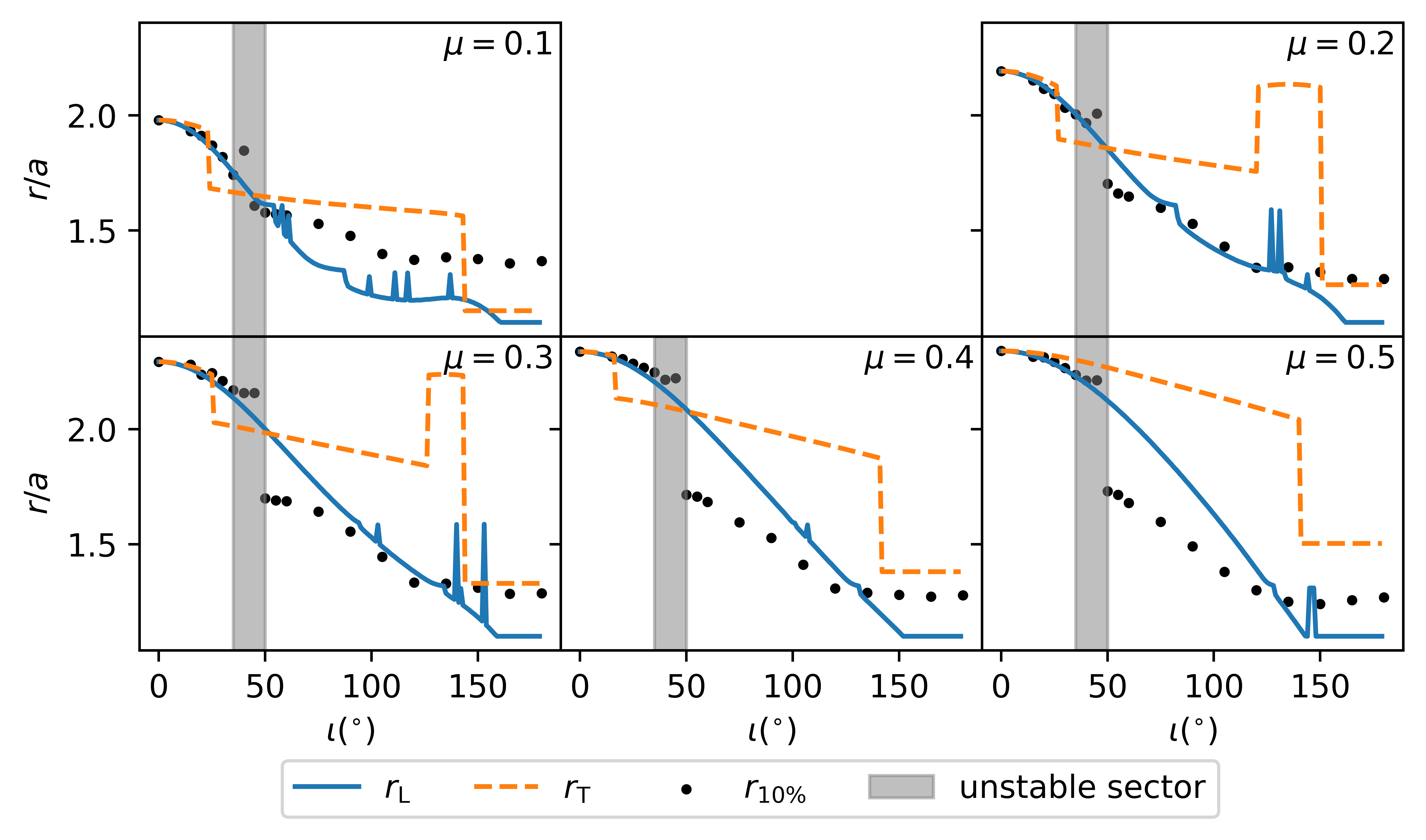}
\caption{The behavior of the gap sizes listed in Table \ref{tab:gap_defs}. The black scatter points denote the numerical results, $r_{10\%}$, with the gray region denoting the unstable sector specified in Sec.~\ref{sec:results}. The solid blue line denotes the gap size $r_{\rm L}$ from the orbital instability picture and the dashed orange line refers to the gap size $r_{\rm T}$ in the resonant torque picture. }
\label{Fig:gapsizecomparison}
\end{figure}

As with Paper I, we wish to compare the two gap opening mechanisms, but with the added dependence on inclination angle. 
We emphasize as with Paper I that we wish to compare the trends in the gap size as opposed the exact values at any given mass ratio or inclination, given the flexibility in each definition of gap size. We normalize the numerical and analytical results to clearly identify the trends they make when we overlay them graphically. In this case, we define the quantities $\bar{r}_\mathrm{X}$ such that,
\begin{equation}
    \bar{r}_\mathrm{X} = \frac{r_\mathrm{L}}{r_\mathrm{X}}\bigg\rvert_{\mu = 0.5, \iota = 0} r_\mathrm{X},
\end{equation}
where the subscript $\mathrm{X}$ denotes the subscript of the relevant radius from the numerical or analytical computations, e.g., $r_\mathrm{10\%}$ as defined in Section \ref{sec:results}. The vertical line denotes that the values are taken at the mass ratio $\mu = 0.5$ and inclination $\iota = 0$, so that the gap size as defined by the instability timescale and the numerical/analytical measure of gap size are normalized to agree for equal masses and a coplanar disk-binary system. Finally, we summarize all the definitions of the gap size in Table \ref{tab:gap_defs}.  We also plot the results of the three different definitions in Figure \ref{Fig:gapsizecomparison}. The black scatter plot data indicates the behavior of the numerical gap size $r_{10\%}$ over inclination, with each plot corresponding to a different mass ratio, and the unstable sector again highlighted as the gray region. The orange dashed line shows the gap size $r_{\rm T}$ derived from the resonant torque picture and the blue solid line shows the gap size $r_{\rm L}$ derived from the orbital stability picture. 
We note that at all mass ratios, the behavior of $r_{10\%}$ shows better agreement with the orbital stability picture than the resonant torque picture. This is particularly true at lower mass ratios where the instability picture predicts gap sizes much closer to the numerical values at inclinations beyond the step function drop. Additionally, at some mass ratios and inclinations, we see an increase in $r_{\rm T}$ due to the dominance of non-circular resonances which is not reflected in the numerical data. 
However, we also observe that $r_{\rm L}$ does not display the step function-like drop in the gap size while $r_{\rm T}$ does. That being said, the step function behavior of $r_{\rm T}$ does not appear to match the step function behavior of $r_{10\%}$. For instance, we note that the critical inclination where the numerical results for $r_{10\%}$ drop does not depend on mass ratio while the size of the drop is larger as we approach the equal mass case. However, in the resonant torquing picture, the drop in gap size $r_{\rm T}$ happens at lower inclinations as we approach the equal mass case, which has no step function like behavior at all, because the drop in $r_{\rm T}$ is due to the transition of the gap opening resonance from the $(1,1)$ Lindblad resonance to the $(2,2)$ Lindblad resonance. On the other hand, the drop in the numerical results is due to the fact that in two different ranges of inclinations, the quasi steady state is arrived at differently. It is unclear exactly why this is the case in our 2-D simulations, and these inclinations are high enough that role of the gravitational force from the binaries directed out of the disk plane may need to be included to gain a complete picture; 3-D simulations could therefore provide a better understanding of how the quasi-steady state is approached in this regime. Additionally, while the orbital stability picture seems to provide better agreement than the resonant torque picture at higher mass ratios (closer to equal mass), the effect of vertical instabilities will need to be incorporated at higher mass ratios where the strength of the gravitational force is higher for larger inclinations. 
Overall, the results of this paper indicate that the formation and maintenance of the gap at binary timescales through the propagation of orbital instabilities appears to be the dominant mechanism over the parameter space of mass ratios and inclinations at the viscosities we consider.

\begin{deluxetable}{c c c}
\tablecaption{Summary of all gap size definitions. \label{tab:gap_defs}}
\tablehead{ Label & Definition & Defined}
\startdata
$r_{10\%}$ & Radius where surface density is $10\%$ of the final maximum & Table \ref{tab:1}\\
$r_{\rm L}$ & Radius where the Lyapunov timescale is equal to the binary orbital period & Equation \eqref{rl}  \\ 
$r_{\rm T}$ & Radius where the epicyclic amplitude reaches the gap-opening resonance &  Equation \eqref{eq:rT}\\
\enddata
\end{deluxetable}

\section{Conclusions}\label{sec:conc}

We have studied black hole binary-disk systems within Newtonian gravity and Newtonian hydrodynamics. We have considered configurations with different mass ratios of the binary and two-dimensional, non-self-gravitating, viscous, locally isothermal disks inclined with respect to the binary's orbital plane. Black hole binaries are modeled by point masses which move on fixed circular orbits.

We investigated the influence of mass ratio and inclination angle on the disk's density distribution. We focused on the location of three radii which give a general description of the matter distribution in the disk, namely: $r_\mathrm{dT}$ --- the radius where the dynamical torque density equals viscous torque density, $r_\mathrm{max}$ --- the radius where the density reaches its maximum, and $r_{10\%}$ --- the radius where the density reaches $10\%$ of the final density maximum. Analyzing these radii together with the general density distributions, we present a broad perspective on the matter density distribution as a function of our model parameters.

Finally, we investigated the effect of different theoretical explanations for the opening and maintenance of the circumbinary gap. In Paper I, we noted that the propagation of orbital instabilities over the timescale of the binary period provides a markedly better estimate of the gap size. In this paper we have found that we can reasonably extend this explanation to the case of inclined binary-disk systems. The theoretical explanations and numerical studies approximate the circumbinary disk as a two-dimensional plane; however, a three-dimensional treatment of the disk may be able to clarify the nature of the ``unstable sector'' of inclination angles, where the disk never reaches a quasi-steady state. We leave this for future study.  

Our explanation for the opening of the circumbinary gap over orbital timescales also motivates future studies of this nature with two other model parameters: the \textit{binary orbital eccentricity}, which has been shown to grow due to the back-reaction of disk dynamics onto the binary (see \cite{2023MNRAS.522.2707S}) or trigger Kozai-Lidov oscillations in highly inclined disks (see \cite{2017ApJ...835L..29F}); and the \textit{binary orbital precession} which is seen in binary black hole systems due to spin-orbit interactions \citep{2014grav.book.....P}.  Future studies of this type, along with the physics discussed in this paper, will have broad implications for these astrophysical systems, which should be regularly observed by gravitational-wave and electromagnetic observatories in the coming decades.

\bibliography{paper2}{}
\bibliographystyle{aasjournal}

\end{document}